\documentclass[12pt]{JHEP3}

\usepackage{epsfig,latexsym,amssymb,cite}
\usepackage{amsmath}
\usepackage{graphicx}
\usepackage{bm}

\newcommand{\be}{\begin{equation}}
\newcommand{\ee}{\end{equation}}
\newcommand{\bea}{\begin{eqnarray}}
\newcommand{\eea}{\end{eqnarray}}
\newcommand{\bem}{\begin{multline}}
\newcommand{\eem}{\end{multline}}
\newcommand{\beg}{\begin{gather}}
\newcommand{\eeg}{\end{gather}}

\newcommand{\as}{\alpha_s}

\def\eq#1{{Eq.~(\ref{#1})}}
\def\fig#1{{Fig.~\ref{#1}}}
\newcommand{\ben}{\begin{eqnarray*}}
\newcommand{\een}{\end{eqnarray*}}
\newcommand{\un}[1]{\underline{#1}}

\graphicspath{{Figs/}}

\title{R-Current DIS on a Shock Wave: Beyond the Eikonal Approximation}

\author{Yuri V. Kovchegov \\
\vspace{0.1in}

Department of Physics, The Ohio State University, Columbus,
OH 43210, USA \\~~\\ 

E-mail address: \email{yuri@mps.ohio-state.edu}

\vspace{0.1in}
}

\date{April 2010}

\abstract{We find the DIS structure functions at strong coupling by
  calculating $R$-current correlators on a finite-size shock wave
  using AdS/CFT correspondence. We improve on the existing results in
  the literature by going beyond the eikonal approximation for the two
  lowest orders in graviton exchanges. We argue that since the eikonal
  approximation at strong coupling resums integer powers of $1/x$
  (with $x$ the Bjorken-$x$ variable), the non-eikonal corrections
  bringing in positive integer powers of $x$ can not be neglected in
  the small-$x$ limit, as the non-eikonal order-$x$ correction to the
  $(n+1)$st term in the eikonal series is of the same order in $x$ as
  the $n$th eikonal term in that series.  We demonstrate that, in
  qualitative agreement with the earlier DIS analysis based on
  calculation of the expectation value of the Wilson loop in the shock
  wave background using AdS/CFT, after inclusion of non-eikonal
  corrections DIS structure functions are described by two momentum
  scales: $Q_1^2 \sim \Lambda^2 \, A^{1/3}/x$ and $Q_2^2 \sim
  \Lambda^2 \, A^{2/3}$, where $\Lambda$ is the typical transverse
  momentum in the shock wave and $A$ is the atomic number if the shock
  wave represents a nucleus. We discuss possible physical meanings of
  the scales $Q_1$ and $Q_2$.}


\preprint{}

\begin{document}


\section{Introduction}

Over the past two decades there have been significant progress in our
theoretical understanding of the physics of parton saturation/Color
Glass Condensate (CGC) \cite{Gribov:1984tu, Mueller:1986wy,
  Mueller:1994rr, Mueller:1994jq, Mueller:1995gb, McLerran:1993ka,
  McLerran:1993ni, McLerran:1994vd, Kovchegov:1996ty,
  Kovchegov:1997pc, Jalilian-Marian:1997xn, Jalilian-Marian:1997jx,
  Jalilian-Marian:1997gr, Jalilian-Marian:1997dw,
  Jalilian-Marian:1998cb, Kovner:2000pt, Weigert:2000gi, Iancu:2000hn,
  Ferreiro:2001qy, Kovchegov:1999yj, Kovchegov:1999ua,
  Balitsky:1996ub, Balitsky:1997mk, Balitsky:1998ya, Iancu:2003xm,
  Weigert:2005us, Jalilian-Marian:2005jf}. The
Jalilian-Marian--Iancu--McLerran--Weigert--Leonidov--Kovner (JIMWLK)
\cite{Jalilian-Marian:1997jx, Jalilian-Marian:1997gr,
  Jalilian-Marian:1997dw, Jalilian-Marian:1998cb, Kovner:2000pt,
  Weigert:2000gi, Iancu:2000hn, Ferreiro:2001qy} and
Balitsky--Kovchegov (BK) \cite{Balitsky:1996ub,
  Balitsky:1997mk,Balitsky:1998ya,Kovchegov:1999yj, Kovchegov:1999ua}
evolution equations have been constructed which unitarize the linear
Balitsky--Fadin--Kuraev--Lipatov (BFKL) \cite{Kuraev:1977fs,Bal-Lip}
evolution equation for DIS on a nucleus. The phenomenological
successes of the CGC physics in describing the data from both the Deep
Inelastic Scattering (DIS) experiments at HERA
\cite{Gotsman:2002yy,Goncalves:2006yt,Golec-Biernat:1999qd,Iancu:2003ge}
and from heavy ion collision experiments at RHIC
\cite{Kharzeev:2000ph,Kharzeev:2001yq,Kharzeev:2004yx,Albacete:2007sm}
allows one to believe that CGC physics correctly captures some of the
main features of QCD dynamics in high energy scattering.

In recent years a number of research efforts have been aimed at
sharpening the quantitative predictive power of CGC/saturation
physics. Running coupling corrections to JIMWLK and BK evolution
equations have been calculated in
\cite{Kovchegov:2006vj,Balitsky:2006wa,Kovchegov:2006wf,Albacete:2007yr}
and led to a marked improvement in the agreement between CGC
predictions and the experimental data
\cite{Albacete:2007sm,Albacete:2009fh}.  Subleading-$N_c$ corrections
to BK evolution were analyzed in \cite{Kovchegov:2008mk} and were
found to be very small, though some DIS observables were shown to be
sensitive to the difference in \cite{Marquet:2010cf}.  Next-to-leading
logarithmic (NLO) corrections to BK evolution equation have been
calculated in \cite{Balitsky:2008zz,Balitsky:2010jf}: the corrections
were found to be in agreement with the NLO BFKL calculation of
\cite{Fadin:1998py,Ciafaloni:1998gs} and, therefore, numerically large
for linear evolution. While it is not clear whether NLO corrections
are large in the solution of the full non-linear NLO BK equation,
since such solution is yet to be obtained, it is important to estimate
the size of higher order corrections to the JIMWLK and BK evolution
equations beyond the running coupling corrections found in
\cite{Kovchegov:2006vj,Balitsky:2006wa,Kovchegov:2006wf,Albacete:2007yr}.
The assessment of the size of higher order correction may happen by
performing explicit higher order calculations of the BK kernel
obtaining a (presumably numerical) solution of the BK equation at each
order.

Alternatively, to estimate the size of higher-order corrections in the
extreme large-coupling limit, one may use the Anti-de Sitter
space/conformal field theory (AdS/CFT) correspondence
\cite{Maldacena:1997re,Gubser:1998bc,Witten:1998qj,Aharony:1999ti} to
study DIS. Indeed AdS/CFT correspondence is a duality between ${\cal
  N}=4$ super Yang--Mills (SYM) theory and type-IIB string theory, and
as such does not apply directly to QCD. Still since ${\cal N}=4$ SYM
theory is a QCD-like gauge theory, i.e., it contains gluodynamics as a
part of the theory, there is hope that many of its qualitative
features and, possibly, some quantitative ones would apply to QCD.
Certainly, on the perturbative side, the leading-order (LO)
(pure-glue) BFKL equation is identical in QCD and in ${\cal N}=4$ SYM,
with many similarities at NLO \cite{Balitsky:2009yp} as well.

High energy scattering in general, and DIS in particular, in the
context of AdS/CFT correspondence has been studied by many groups
\cite{Janik:1999zk,Polchinski:2000uf,Polchinski:2002jw,BallonBayona:2007qr,BallonBayona:2007rs,Cornalba:2008sp,Cornalba:2009ax,Cornalba:2007zb,Cornalba:2010vk,Brower:2006ea,Brower:2007qh,Hatta:2007he,Hatta:2007cs,Albacete:2008ze,Kovchegov:2009yj,Levin:2008vj,Mueller:2008bt,Avsar:2009xf,Brodsky:2003px,Marquet:2009ca}.
In those works the pomeron intercept at large 't Hooft coupling
$\lambda = g^2 \, N_c$ has been calculated
\cite{Janik:1999zk,Brower:2006ea,Hatta:2007he,Albacete:2008ze}, though
some disagreement still exists about its precise value
\cite{Giordano:2010yb,Taliotis:2009ne}. Another important quantity for
elucidating higher-order corrections to saturation/CGC physics is the
saturation scale $Q_s$, a momentum scale below which, in the
perturbative framework, the non-linear saturation effects become
important \cite{Iancu:2003xm, Weigert:2005us, Jalilian-Marian:2005jf}.
In CGC approaches based on LO BK or JIMWLK evolution the saturation
scale grows as an inverse power of Bjorken-$x$ variable and as a power
of the nuclear atomic number $A$ for DIS on a nucleus, $Q_s^2 \sim
A^{1/3} \, (1/x)^{\text{const} \, \as}$, with $\as$ the strong
coupling constant.  In the AdS/CFT framework the saturation scale has
been calculated in \cite{Hatta:2007he,Hatta:2007cs} for DIS on an
infinite thermal medium with the result that $Q_s^2 \sim 1/x^2$ at
large $\lambda$. In \cite{Hatta:2007he,Hatta:2007cs}, following
\cite{Polchinski:2000uf,Polchinski:2002jw}, electromagnetic current of
the standard model was replaced by the $R$-current in ${\cal N}=4$ SYM
theory. The hadronic tensor of DIS was then replaced by a correlator
of two $R$-currents, which was calculated at large 't Hooft coupling
using the methods of AdS/CFT correspondence. Generalization of the
method of \cite{Hatta:2007he,Hatta:2007cs} to the case of DIS on a
finite-size medium (modeling a proton or a nucleus) was done in
\cite{Mueller:2008bt,Avsar:2009xf}. The saturation scale obtained in
\cite{Mueller:2008bt,Avsar:2009xf} scaled as $Q_s^2 \sim A^{1/3}/x$.

An alternative approach to DIS was suggested in
\cite{Albacete:2008ze}: in QCD it is well-known that DIS at small-$x$
can be viewed as virtual photon splitting into a quark--anti-quark
pair with the pair interacting with the proton/nucleus
\cite{Kovchegov:1999yj,Jalilian-Marian:2005jf}. Since only the
interaction of the quark dipole with the proton/nucleus is described
by strong interactions, only this part of the DIS cross section can be
strongly coupled and should be modeled using AdS/CFT. In
\cite{Albacete:2008ze} the forward scattering amplitude of a dipole on
a nucleus has been calculated modeling the ultrarelativistic nucleus
by a shock wave in AdS$_5$. Expectation value of the corresponding
Wilson loop in the shock wave background was then calculated using the
AdS/CFT prescription \cite{Maldacena:1998im}. The dipole scattering
amplitude obtained in \cite{Albacete:2008ze} allowed for successful
descriptions of some of the HERA DIS data
\cite{Kovchegov:2009yj,Betemps:2010ij}, albeit in a limited region of
small photon virtuality $Q^2$ where QCD coupling constant should be
large. While the shock wave considered in \cite{Albacete:2008ze} had a
finite longitudinal extent, as we will show below the results of
\cite{Albacete:2008ze} can be easily generalized to an infinite-size
shock wave, giving the saturation scale $Q_s^2 \sim 1/x^2$, in
agreement with \cite{Hatta:2007he,Hatta:2007cs}. However, the
saturation scale for the interesting and realistic case of a
finite-size shock wave found in \cite{Albacete:2008ze} scales as $Q_s
\sim A^{1/3} \, (1/x)^0 \sim A^{1/3}$, in disagreement with the
results of \cite{Mueller:2008bt,Avsar:2009xf,Levin:2008vj} (though in
apparent agreement with \cite{Dominguez:2008vd}, where a similar
method of inserting a fundamental string in the bulk was used, though
for the purpose of jet quenching studies).

The goal of the present paper is to attempt to reconcile the results
of \cite{Albacete:2008ze} with that of
\cite{Mueller:2008bt,Avsar:2009xf,Levin:2008vj} and/or to elucidate
the origin of the discrepancy. We will try to perform $R$-current DIS
calculation without employing the eikonal approximation used in
\cite{Mueller:2008bt,Avsar:2009xf,Levin:2008vj}.  Our motivation is
the following. The eikonal series of graviton exchanges in AdS/CFT
sums up powers of $1/x$ on the gauge theory side.  If $x$ is small,
this is a series in powers of a large number $1/x$, and, as such, is
susceptible to corrections. Namely, order-$x$ non-eikonal correction
to the $(n+1)$st term in the series is $(1/x)^{n+1} \times x =
(1/x)^n$, i.e., it is of the same order as the $n$th term in the
series. Since the coefficients in the eikonal series are functions of
$Q^2$, the condition of non-eikonal corrections being small translates
into a bound on $Q^2$. Below we will show that the eikonal approach of
\cite{Mueller:2008bt,Avsar:2009xf,Levin:2008vj} is valid only for $Q^2
\gtrsim (Q_1)^2$ with $(Q_1)^2 \sim A^{1/3}/x$ the candidate for the
saturation scale found in
\cite{Mueller:2008bt,Avsar:2009xf,Levin:2008vj}. The breakdown of
eikonal approximation is due to the presence of another scale in the
problem, $Q_2 \sim A^{1/3}$, which corresponds to the candidate for
the saturation scale found in \cite{Albacete:2008ze}. Our conclusion
is that $R$-current DIS is a two-scale problem and that the exact
solution of the problem should determine which of the scales $Q_1$ and
$Q_2$ is the saturation scale in strong-coupling DIS.

The paper is structured as follows. We start in Sec. (\ref{setup}) by
defining all the main concepts and quantities used in the calculation.
In Sec. (\ref{GE}) we construct general exact expressions for the
hadronic tensor modeled in AdS/CFT. The expressions for two
independent components of the hadronic tensor are given in
Eqs.~(\ref{Pij3}) and (\ref{P++2}) below. As these expressions appear
to be too complicated to be evaluated precisely analytically, here we
first evaluate them using the eikonal approximation of
\cite{Mueller:2008bt,Levin:2008vj,Albacete:2009ji,Avsar:2009xf} in
Sec. \ref{sec:eik}. In the process we find the applicability region of
the eikonal approximation: $Q^2 \gtrsim (Q_1)^2$ (see Sec.
\ref{appl}). To solidify this conclusion we evaluate Eqs.~(\ref{Pij3})
and (\ref{P++2}) exactly order-by-order in graviton exchanges to the
first non-trivial order in Sec. \ref{sec:beyond} and show explicitly
when the non-eikonal corrections become comparable to the eikonal
terms. We summarize the results of our calculations in Sec.
\ref{sec:sum}. Finally, in Sec. \ref{MS} we conclude by outlining some
of the possible physical interpretations of the scales $Q_1$ and
$Q_2$.


\section{General Setup}
\label{setup}

Our goal is to model DIS on a shock wave at strong coupling. For
simplicity we will consider shock waves without transverse coordinate
dependence in their profile. In \cite{Janik:2005zt}, using the
holographic renormalization \cite{deHaro:2000xn}, the geometry in
AdS$_5$ dual to a relativistic nucleus in the boundary theory was
suggested to be given by the following metric
\begin{align}\label{nuc}
  ds^2 \, = \, \frac{L^2}{z^2} \, \left\{ -2 \, dx^+ \, dx^- + \Phi
    (x^-) \, z^4 \, d x^{- \, 2} + d x_\perp^2 + d z^2 \right\}. 
\end{align}
Here $d x_\perp^2 = (d x^1 )^2 + (d x^2)^2$ is the transverse metric
and $x^\pm = (x^0 \pm x^3) / \sqrt{2}$ where $x^3$ is the collision
axis. $L$ is the radius of S$_5$ and $z$ is the coordinate describing
the 5th dimension with the boundary of AdS$_5$ at $z=0$. \eq{nuc} is
the solution of Einstein equations in AdS$_5$ for $\Phi (x^-)$ being
an arbitrary function of $x^-$. 

According to the AdS/CFT prescription \cite{deHaro:2000xn} the
energy-momentum tensor in the boundary gauge theory dual to the metric
(\ref{nuc}) has only one non-vanishing component:
\begin{align}
  T_{--} (x^-) \, = \, \frac{N_c^2}{2 \, \pi^2} \, \Phi (x^-).
\end{align}
Thus different functions $\Phi (x^-)$ correspond to different
longitudinal profiles of the nuclear energy-momentum tensor.

Here we take a shock wave made of homogeneous matter with a finite
longitudinal extent \cite{Albacete:2008vs,Albacete:2009ji}
\begin{align}\label{Phi}
  \Phi (x^-) = \frac{\mu}{a} \, \theta (x^-) \, \theta (a-x^-). 
\end{align}
While this is a simple ansatz, it appears to be quite realistic for a
large ultrarelativistic nucleus. Indeed transverse coordinate
dependence is neglected in \eq{Phi}, but since the relevant transverse
distance scales for a DIS process are much shorter than the length of
a typical variation of the nuclear profile in the transverse
direction, we believe neglecting transverse dependence in \eq{Phi}
does not affect the physics in a qualitative way and is a good first
approximation for studying DIS in AdS/CFT, which can be easily
improved upon later.

The parameter $\mu$ is related to the large light-cone momentum of the
nucleons in the nucleus $p^+$, the the atomic number $A$ and the
typical transverse momentum scale $\Lambda$ by \cite{Albacete:2008vs}
\begin{align}\label{mu}
  \mu \propto p^+ \, \Lambda^2 \, A^{1/3}.
\end{align}
The longitudinal width of the nucleus is Lorentz-contracted and is
given approximately by \cite{Albacete:2008vs}
\begin{align}\label{a}
  a \sim \frac{A^{1/3}}{p^+}.
\end{align}

Following \cite{Hatta:2007he,Hatta:2007cs} we will model
electromagnetic current in DIS by an $R$-current $J_\mu (x)$, which is
a conserved current corresponding to a $U(1)$ subgroup of the $SU(4)$
$R$-symmetry of the ${\cal N} =4$ SYM theory in four dimensions. The
current $J_\mu (x)$ can be written in terms of the scalar, spinor and
vector fields of the ${\cal N} =4$ SYM theory. For a systematic and
pedagogical definition of the $R$-current we refer the reader to
\cite{Bartels:2008zy,CaronHuot:2006te}.

We want to calculate the retarded $R$-current correlator
\cite{Hatta:2007he,Hatta:2007cs,McLerran:1998nk,Avsar:2009xf,Son:2002sd,Son:2007vk,Policastro:2002se}
\begin{align}\label{Pdef}
  \Pi^{\mu\nu} (q) \, = \, \frac{\Lambda^2}{a} \, i \, \int d^2
  x_\perp \, d^2 y_\perp \, d x^- \, d y^- \, d (x^+ - y^+) \, e^{- i
    \, q \, \cdot \, (x-y)} \, \theta (x^0 - y^0) \, \langle p| \left[
    J^\mu (x) , J^\nu (y) \right] | p \rangle
\end{align}
where all integrals run from $-\infty$ to $+\infty$. Our metric in
4-dimensions is $\eta_{\mu\nu} = diag (-1, +1, +1, +1)$. By
$|p\rangle$ we denote the proton or nucleus state which is modeled in
AdS/CFT by the shock wave (\ref{nuc}). The essential ingredient of
\eq{Pdef} is the retarded Green function in the coordinate space
\begin{align}\label{retG}
  \Pi^{\mu\nu} (x,y) \, = \, i \, \theta (x^0 - y^0) \, \langle p|
  \left[ J^\mu (x) , J^\nu (y) \right] |p \rangle.
\end{align}
Note that, as usual in DIS, current conservation and Lorentz
symmetries demand that $\Pi^{\mu\nu} (q)$ can be written in the
following standard form
\begin{align}\label{strf_def}
  \Pi^{\mu\nu} (q) \, = \, \left( \eta^{\mu\nu} - \frac{q^\mu \,
      q^\nu}{q^2} \right) \, \Pi_1 (x, Q^2) + \left( p^{\mu} - \frac{p
      \cdot q}{q^2} \, q^\mu \right) \, \left( p^{\nu} - \frac{p \cdot
      q}{q^2} \, q^\nu \right) \, \Pi_2 (x, Q^2)
\end{align}
where $p$ is the momentum of (a nucleon in) the shock wave, 
\begin{align}
  Q^2 = q^2,
\end{align}
and the Bjorken-$x$ variable is
\begin{align}\label{xBj}
  x = \frac{Q^2}{- 2 \, p \cdot q}.
\end{align}
(Note again our $\eta_{\mu\nu} = diag (-1, +1, +1, +1)$ metric
convention.) The imaginary part of the correlator $\Pi^{\mu\nu} (q)$
is proportional to the DIS hadronic tensor: nevertheless, for brevity,
we will refer to $\Pi^{\mu\nu} (q)$ itself as hadronic tensor.

To calculate the retarded Green function at strong 't Hooft coupling
using AdS/CFT correspondence one makes use of the fact that the
$R$-current $J^\mu$ is dual to a Maxwell gauge field in the bulk
\cite{Freedman:1998tz,Chalmers:1998xr,Policastro:2002se,Son:2002sd,Son:2007vk}.
The action of the Maxwell gauge field in empty AdS$_5$ space and in
the space described by the metric (\ref{nuc}) is
\begin{align}\label{Mact}
  S_{\text{Maxwell}} \, = \, - \frac{N_c^2}{64 \, \pi^2 \, L} \, \int
  d^5 x \, \sqrt{-g} \, F_{MN} \, F^{MN} \, = \, - \frac{N_c^2 \,
    L^4}{64 \, \pi^2} \, \int d^4 x \, \frac{dz}{z^5} \, F_{MN} \,
  F^{MN}.
\end{align}
Here and throughout the paper indices $M,N$ run from $0$ to $4$, while
$\mu,\nu$ run from $0$ to $3$.

Classical sourceless Maxwell equations in the curved background read
\begin{align}\label{Maxwell}
  \partial_M \left[ \sqrt{-g} \, g^{MN} \, g^{RS} \, F_{NS} \right] \,
  = \, 0.
\end{align}
In AdS$_5$ the classical Maxwell action can be written with the help
of \eq{Maxwell} as \cite{Hatta:2007cs}
\begin{align}\label{Mcl}
  S^{cl}_{\text{Maxwell}} \, = \, - \frac{N_c^2}{32 \, \pi^2} \, \int
  d^4 {\tilde x} \, \left[ \frac{1}{z} \left( A_+ \partial_z A_- + A_-
      \partial_z A_+ - A_i \partial_z A_i \right) \right]\bigg|_{z=0}
\end{align}
with $i=1,2$ denoting transverse spatial dimensions and with summation
assumed over repeated indices. In arriving at \eq{Mcl} we have made
use of $A_z =0$ gauge, which we will employ from now on.

In the background of the metric (\ref{nuc}) with the shock wave
profile (\ref{Phi}) Maxwell equations become (labeled by component)
\begin{subequations}\label{Maxwell_all}
\begin{align}
  & (+) \hspace*{0.1in} \left[ z \, \partial_+ \, \partial_- +
    \partial_z - z \, \partial_z^2 \right] \, A_- (x^+, x^-, z) - z \,
  \partial_-^2 \, A_+ (x^+, x^-, z) \notag \\ & \hspace*{5cm} \, = \, \frac{\mu}{a}
  \, z^4 \, \theta (x^-) \, \theta
  (a-x^-) \, \left[ 3 \, \partial_z + z \, \partial_z^2 \right] \, A_+ (x^+, x^-, z), \\
  & (-) \hspace*{0.1in} \left[ z \, \partial_+ \, \partial_- +
    \partial_z - z \, \partial_z^2 \right] \, A_+ (x^+, x^-, z)
  \, = z \, \, \partial_+^2 \, A_- (x^+, x^-, z), \label{M-} \\
  & (\bot) \hspace*{0.1in} \left[ 2 \, \partial_+ \, \partial_- +
    \frac{1}{z} \, \partial_z - \partial_z^2 \right] \, A_i (x^+, x^-,
  z) \, = \, - \frac{\mu}{a} \, z^4 \, \theta (x^-) \, \theta (a-x^-)
  \,
  \partial_+^2 \, A_i (x^+, x^-, z), \label{Mperp} \\
  & (z) \hspace*{0.1in} \partial_z \left[ \partial_- A_+ (x^+, x^-, z)
    + \partial_+ A_- (x^+, x^-, z) \right] = - \frac{\mu}{a} \, z^4 \,
  \theta (x^-) \, \theta (a-x^-) \, \partial_z \, \partial_+ \, A_+
  (x^+, x^-, z). \label{Mz}
\end{align}
\end{subequations}
In arriving at Eqs.~(\ref{Maxwell_all}) we have assumed that the gauge
field $A_\mu (x,z)$ is independent of the transverse coordinates
${\underline x} = (x^1, x^2)$. The reason for this assumption will be
explained later.

The AdS/CFT prescription for calculating this retarded Green function
(\ref{retG}) is
\cite{Freedman:1998tz,Chalmers:1998xr,Policastro:2002se,Son:2002sd,Son:2007vk}
\begin{align}\label{corr_pr}
  \Pi^{\mu\nu} (x,y) \, = \, \frac{\delta
    S^{cl}_{\text{Maxwell}}}{\delta A^b_\mu (x) \, \delta A^b_\nu (y)}
\end{align}
where $A^b_\mu (x)$ is the value of the classical Maxwell gauge field
at the boundary of AdS$_5$. However, the correlator one would obtain
from \eq{corr_pr} in the background of the metric given by Eqs.
(\ref{nuc}) and (\ref{Phi}) would contain both the vacuum component
and the $\mu$-dependent term due to DIS on a shock wave. Since we are
interested in the latter we need to subtract the vacuum piece.  We
thus write using \eq{Pdef}
\begin{align}\label{Pexpr}
  \Pi^{\mu\nu} (q) \, = \, \frac{\Lambda^2}{a} \int d^2 x_\perp \,
  d^2 y_\perp \, d x^- \, d y^- \, d (x^+ - y^+) \, e^{- i \, q \,
    \cdot \, (x-y)} \, \frac{\delta^2 }{\delta A^b_\mu (x) \, \delta
    A^b_\nu (y)} \left[ S^{cl}_{\text{Maxwell}} (\mu) -
    S^{cl}_{\text{Maxwell}} (0) \right],
\end{align}
where $S^{cl}_{\text{Maxwell}} (\mu)$ is the classical Maxwell field
action in the background of the shock wave metric (\ref{nuc}), while
$S^{cl}_{\text{Maxwell}} (0)$ is the same action in the empty AdS$_5$
background.


\section{$R$-Currents Correlator}


\subsection{General Expression}
\label{GE}


\subsubsection{Transverse Components of the Hadronic Tensor}

While Eqs.~(\ref{Maxwell_all}) are hard to solve exactly, we will look
for the solution perturbatively in $\mu$. We start by concentrating on
the transverse field component $A_i$ which contributes to the
transverse part $\Pi^{ij}$ of the hadronic tensor $\Pi^{\mu\nu}$. Note
that the equation (\ref{Mperp}) for $A_i$ completely decouples from
the rest of Maxwell equations~(\ref{Maxwell_all}): therefore we can
treat $A_i$ as an independent degree of freedom. We write
\begin{align}\label{muexp}
  A_i (x,z) \, = \, A_i^{(0)} (x,z) + A_i^{(1)} (x,z) + A_i^{(2)}
  (x,z) + \ldots
\end{align}
where the term $A_i^{(n)}$ is of the order $\mu^n$.

Start by putting $\mu=0$ (no shock wave) and solving \eq{Mperp} for
$A_i^{(0)} (x,z)$
\begin{align}\label{M0}
  \left[ 2 \, \partial_+ \, \partial_- + \frac{1}{z} \, \partial_z -
    \partial_z^2 \right] \, A_i^{(0)} (x^+, x^-, z) \, = \, 0.
\end{align}
Concentrating on the $z$-dependence of $A_i^{(0)} (x,z)$ we see that
the general solution of this equation can be written as
\begin{align}\label{Aigen}
  A_i^{(0)} (x^+, x^-, z) \, = \, z \, \sqrt{2 \, \partial_+ \,
    \partial_-} \, K_1 (z \, \sqrt{2 \, \partial_+ \, \partial_-}) \, C_1 (x^+, x^-)+
   z \, \sqrt{2 \, \partial_+ \, \partial_-} \, I_1 (z \,
  \sqrt{2 \, \partial_+ \, \partial_-}) \, C_2  (x^+, x^-)
\end{align}
with $C_1$ and $C_2$ some arbitrary functions.  Demanding that our
Maxwell field (and, more importantly, its field strength) grows slower
than $\sim z^1$ as $z \rightarrow \infty$\footnote{This condition
  arises in deriving \eq{Mcl}, where the contribution from $z=\infty$
  can be neglected only if all field components grow slower than $\sim
  z^1$ at large-$z$.} we can discard the second term on the right of
\eq{Aigen} since it would give an exponential divergence at large-$z$
for positive eigenvalues of the operator $2 \, \partial_+ \,
\partial_-$.  We therefore write
\begin{align}\label{Ai0}
  A_i^{(0)} (x^+, x^- ,z) \, = \, z \, \sqrt{2 \, \partial_+ \,
    \partial_-} \, K_1 (z \, \sqrt{2 \, \partial_+ \, \partial_-}) \,
  A_i^b (x^+, x^-)
\end{align}
where $A_i^b (x)$ is the boundary value of the field $A_i^{(0)}
(x,z)$.

We define the Green function of the operator on the left-hand-side of
\eq{Mperp} (while treating it as a differential operator in $z$ only)
by
\begin{align}\label{Green}
  \left[ 2 \, \partial_+ \, \partial_- + \frac{1}{z} \, \partial_z -
    \partial_z^2 \right] \, G (z; z'; \partial_+ \, \partial_-) \, =
  \, z' \, \delta (z-z').
\end{align}
The Green function $G (z; z'; \partial_+ \, \partial_-)$ is itself a
differential operator being dependent on $\partial_+ \, \partial_-$.
Unfortunately \eq{Green} does not uniquely define the Green function
$G$ since we can always shift the Green function by the
right-hand-side of \eq{Aigen} unless we specify the boundary
conditions. As one can see from \eq{Pexpr} our goal is to
differentiate with respect to the boundary values of the Maxwell
field. At the same time, if we solve Maxwell equations
(\ref{Maxwell_all}) order-by-order in $\mu$ the boundary value of the
free field in \eq{Ai0} may be modified by $\mu$-dependent corrections
and may become a $\mu$-dependent function itself. In other words the
boundary value of the field $A_i (x,z)$ would not be $A_i^b (x)$ from
\eq{Ai0}, but instead would contain some explicit $\mu$-dependent
terms added to it: it would then be unclear how to perform the
functional differentiation of \eq{Pexpr}. To avoid this problem it
appears easiest to follow \cite{Avsar:2009xf} and demand that
$A_i^{(n)} (x,z)$ is zero at $z=0$ for all $n \ge 1$. Then the
boundary value of the full field $A_i (x,z)$ would be given by $A_i^b
(x^+, x^-)$ from \eq{Ai0}, making functional differentiation possible.

We therefore demand that the Green function $G (z; z'; \partial_+ \,
\partial_-)$ is $0$ at $z=0$. On top of that we demand that $G (z; z';
\partial_+ \, \partial_-)$ does not diverge exponentially as $z
\rightarrow \infty$.  These two conditions together with \eq{Green}
fix the Green function $G (z; z'; \partial_+ \, \partial_-)$ uniquely.
The Green function can be shown to be equal to \cite{Hatta:2007cs}
\begin{align}\label{GIK}
  G (z; z'; \partial_+ \, \partial_-) \, = \, z \, z' \, I_1 \left(
    z_< \, \sqrt{2 \, \partial_+ \, \partial_-} \right) \, K_1 \left(
    z_> \, \sqrt{2 \, \partial_+ \, \partial_-} \right)
\end{align}
with 
\begin{align}
  z_{> \, (<)} \, = \, \text{max} \, (\text{min}) \, \{ z, z'\}.
\end{align}

In general the convolution of Green function $G (z; z'; \partial_+ \,
\partial_-)$ with other functions may generate inverse powers of
$\partial_+ \, \partial_-$. Requiring causality we will understand
those as denoting the following operations
\begin{align}\label{ints}
  \frac{1}{\partial_{+}} [\ldots](x^+) \, \equiv \,
  \int\limits_{-\infty}^{x^+} \, d x'^+ \, [\ldots](x'^+), \ \ \ 
  \frac{1}{\partial_{-}} [\ldots](x^-) \, \equiv \,
  \int\limits_{-\infty}^{x^-} \, d x'^- \, [\ldots](x'^-).
\end{align}
Let us define one more abbreviated notation: 
\begin{align}\label{Gabr}
  {\hat G}_z (\partial_+ \, \partial_-) \, f(x^+, x^-, z) \, \equiv \,
  \int_0^\infty \, \frac{d z'}{z'} \, G (z; z'; \partial_+ \,
  \partial_-) \, f(x^+, x^-, z')
\end{align}
for an arbitrary function $f(x^+, x^-, z)$. 

With the help of this notation we write the result of the first
iteration of \eq{Mperp} as
\begin{align}
  A_i^{(1)} & \, (x^+, x^-,z) \, = \, - \frac{\mu}{a} \, {\hat G}_z
  (\partial_+ \, \partial_-) \, z^4 \, \theta (x^-) \, \theta (a-x^-)
  \, \partial_+^2 \, A_i^{(0)} (x^+, x^-,z) \notag \\ & = \, -
  \frac{\mu}{a} \, {\hat G}_z (\partial_+ \, \partial_-) \, z^4 \,
  \theta (x^-) \, \theta (a-x^-) \, \partial_+^2 \, z \, \sqrt{2 \,
    \partial_+ \, \partial_-} \, K_1 (z \, \sqrt{2 \, \partial_+ \,
    \partial_-}) \, A_i^b (x^+, x^-),
\end{align}
where in the last step we have used \eq{Ai0}.  Repeating the procedure
several times a general term in the series of \eq{muexp} can be
written as
\begin{align}
  A_i^{(n)} (x^+, x^-,z) \, = \, \left[ - \frac{\mu}{a} \, {\hat G}_z
    (\partial_+ \, \partial_-) \, z^4 \, \theta (x^-) \, \theta
    (a-x^-) \, \partial_+^2 \right]^n \, z \, \sqrt{2 \, \partial_+ \,
    \partial_-} \, K_1 (z \, \sqrt{2 \, \partial_+ \, \partial_-}) \,
  A_i^b (x^+, x^-),
\end{align}
where each differential operator in the square brackets acts on
everything to its right, that is, say, $\partial_-$ in one of the
brackets acts on all $x^-$-dependence in all other brackets to its
right and on $A_i^b (x^+, x^-)$.

The solution of \eq{Mperp} can then be written as an infinite series
\begin{align}\label{Aisol}
  A_i (x^+, x^-,z) \, = \, \sum\limits_{n=0}^\infty \left[ -
    \frac{\mu}{a} \, {\hat G}_z (\partial_+ \, \partial_-) \, z^4 \,
    \theta (x^-) \, \theta (a-x^-) \, \partial_+^2 \right]^n z \,
  \sqrt{2 \, \partial_+ \, \partial_-} \, K_1 (z \, \sqrt{2 \,
    \partial_+ \, \partial_-}) \, A_i^b (x^+, x^-).
\end{align}

In evaluating structure functions using \eq{Pexpr} it will be
important to know the small-$z$ behavior of $A_i (x^+, x^-,z)$.
Expanding \eq{Ai0} in powers of $z$ yields
\begin{align}\label{Ai0z2}
  A_i^{(0)} (x^+, x^- ,z) \, = \, \left\{ 1 + \frac{z^2 \, 2 \,
      \partial_+ \, \partial_-}{4} \, \left[ \ln \left( \frac{z^2 \, 2
          \, \partial_+ \, \partial_- }{4} \right) + 2 \, \gamma -1
    \right] + o\left(z^4 \, \ln z\right) \right\} \, A_i^b (x^+, x^-).
\end{align}
For $A_i^{(n)}$ with $n \ge 1$ the expansion is different. Expanding
the Green function in \eq{GIK} at small-$z$ we write for $n \ge 1$
\begin{align}\label{Ainz2}
  A_i^{(n)} (x^+, x^-,z) \, = \, - z^2 \, \frac{\sqrt{2 \, \partial_+
      \, \partial_-}}{2} \, \frac{\mu}{a} \, \int\limits_0^\infty \, d
  z' \, K_1 (z' \, \sqrt{2 \, \partial_+ \, \partial_-}) \, z'^4 \,
  \theta (x^-) \, \theta (a-x^-) \, \partial_+^2 \notag \\ \times \,
  \left[ - \frac{\mu}{a} \, {\hat G}_{z'} (\partial_+ \, \partial_-)
    \, z'^4 \, \theta (x^-) \, \theta (a-x^-) \, \partial_+^2
  \right]^{n-1} \, z' \, \sqrt{2 \, \partial_+ \, \partial_-} \, K_1
  (z' \, \sqrt{2 \, \partial_+ \, \partial_-}) \, A_i^b (x^+, x^-) +
  o\left(z^4 \right).
\end{align}

We are now almost ready to evaluate $\Pi^{ij} (q)$. Using \eq{Mcl} in
\eq{Pexpr} along with the asymptotics found in Eqs.~(\ref{Ai0z2}) and
(\ref{Ainz2}) we obtain
\begin{align}\label{Pij1}
  \Pi^{ij} (q) \, & = \, \frac{\Lambda^2}{a} \, \frac{N_c^2}{16 \,
    \pi^2} \, \int d^2 x_\perp \, d^2 y_\perp \, d x^- \, d y^- \, d
  (x^+ - y^+) \, e^{- i \, q \, \cdot \, (x-y)} \, \int d^4 {\tilde x} \notag \\
  & \times \, \left[ \frac{1}{z} \frac{\delta A_k^{(0)} ({\tilde x},z
      )}{\delta A_i^b (x) } \, \partial_z \frac{\delta A_k ({\tilde
        x},z )}{\delta A_j^b (y) } - \frac{1}{z} \frac{\delta
      A_k^{(0)} ({\tilde x},z )}{\delta A_i^b (x) } \, \partial_z
    \frac{\delta A_k^{(0)} ({\tilde x},z )}{\delta A_j^b (y) }
  \right]\Bigg|_{z=0},
\end{align}
where we made use of the fact that $\Pi^{ij} (q)$ is an even function
of $q^\mu$ (see \eq{strf_def}). The sum over $k$ runs over $k=1,2$.

Just like \cite{Avsar:2009xf} we will work in a frame with ${\un q}
=0$. This is the reason we have neglected transverse coordinate
dependence of the classical Maxwell fields throughout the discussion.
The argument is as follows: since the metric tensor in \eq{nuc} is
independent of transverse coordinates, putting transverse coordinate
dependence back into Maxwell equations (\ref{Maxwell_all}) would only
generate some differential operators $\partial_{x_\perp}$ in its
solution (\ref{Aisol}). After substituting \eq{Aisol} into \eq{Pij1},
those differential operators become $\partial_{{\tilde x}_\perp}$
acting either on $\delta^2 ({\un x} - {\un {\tilde x}})$ or on
$\delta^2 ({\un y} - {\un {\tilde x}})$. We can therefore replace
$\partial_{{\tilde x}_\perp} \rightarrow - \partial_{x_\perp}$ (or
$\partial_{{\tilde x}_\perp} \rightarrow - \partial_{y_\perp}$) and
integrate by parts, such that in the end we would get
$\partial_{{\tilde x}_\perp} \rightarrow - \partial_{x_\perp}
\rightarrow - i \, q_\perp$ (or $\partial_{{\tilde x}_\perp}
\rightarrow - \partial_{y_\perp} \rightarrow i \, q_\perp$). Hence
these transverse coordinate derivatives vanish in the $q_\perp =0$
limit, leaving only two transverse delta-functions $\delta^2 ({\un x}
- {\un {\tilde x}}) \, \delta^2 ({\un y} - {\un {\tilde x}})$, which
eliminate ${\tilde x}_\perp$ and $y_\perp$ integrals. The remaining
$x_\perp$ integral is strictly-speaking infinite, but we assume that
the nucleus has a large but finite transverse extent and replace
\begin{align}\label{tr_repl}
  \int d^2 x_\perp \, \rightarrow \, S_\perp
\end{align}
where $S_\perp$ is the transverse area of the nucleus. We assume that
as long as the nucleus is large enough in transverse direction, much
larger than the typical relevant distance scale for DIS, the DIS
process would most of the time be insensitive to the edge effects
justifying the approximation.  This is done in complete analogy with
perturbative DIS calculations
\cite{Kovchegov:1999yj,Kovchegov:1999ua}. We conclude that at $q_\perp
=0$ all transverse integrals simply disappear, leaving only the factor
of $S_\perp$ from \eq{tr_repl}.

Using Eqs.~(\ref{Ai0z2}) and (\ref{Ainz2}) we can find the functional
derivatives needed in \eq{Pij1}:
\begin{align}\label{Ai0der}
  \frac{\delta A_k^{(0)} (x^+, x^-, z )}{\delta A_i^b (y^+, y^-) } \,
  = \, \delta^{ik} \, \delta (x^+ - y^+) \, \delta (x^- - y^-) +
  o\left(z^2 \, \ln z\right)
\end{align}
and for $n \ge 1$
\begin{align}\label{Ainder}
  \frac{\delta A_k^{(n)}(x^+, x^-, z )}{\delta A_i^b (y^+, y^-) } \, &
  = \, - \delta^{ik} \, z^2 \, \frac{\sqrt{2 \, \partial_+ \,
      \partial_-}}{2} \, \frac{\mu}{a} \, \int\limits_0^\infty \, d z'
  \, K_1 (z' \, \sqrt{2 \, \partial_+ \, \partial_-}) \, z'^4 \,
  \theta (x^-) \, \theta (a-x^-) \, \partial_+^2 \notag \\ & \times \,
  \left[ - \frac{\mu}{a} \, {\hat G}_{z'} (\partial_+ \, \partial_-)
    \, z'^4 \, \theta (x^-) \, \theta (a-x^-) \, \partial_+^2
  \right]^{n-1} \, z' \, \sqrt{2 \, \partial_+ \, \partial_-} \, K_1
  (z' \, \sqrt{2 \, \partial_+ \, \partial_-}) \notag \\ & \times \,
  \delta (x^+ - y^+) \, \delta (x^- - y^-) + o\left(z^4 \right).
\end{align}
Here $\delta^{ik}$ is the Kronecker delta. 

Substituting Eqs. (\ref{Ai0der}) and (\ref{Ainder}) into \eq{Pij1} and
employing the arguments which led to \eq{tr_repl} we therefore get
\begin{align}\label{Pij2}
  \Pi^{ij} & \, (q^+, q^-, {\un q}=0) \, = \, - \frac{\Lambda^2 \,
    S_\perp}{a} \, \frac{N_c^2}{8 \, \pi^2} \, \frac{\mu}{2 \, a} \,
  \delta^{ij} \, \int\limits_{-\infty}^\infty d x^- \, d y^- \, d (x^+
  - y^+) \, e^{i \, q^+ \,(x^- - y^-) + i \, q^- \,(x^+ - y^+)} \,
  \sqrt{2 \, \partial_+ \, \partial_-} \notag \\ & \times \,
  \int\limits_0^\infty \, d z \, K_1 (z \, \sqrt{2 \, \partial_+ \,
    \partial_-}) \, z^4 \, \theta (x^-) \, \theta (a-x^-) \,
  \partial_+^2 \, \sum\limits_{n=1}^\infty \, \left[ - \frac{\mu}{a}
    \, {\hat G}_{z} (\partial_+ \, \partial_-) \, z^4 \, \theta (x^-)
    \, \theta (a-x^-) \, \partial_+^2 \right]^{n-1} \notag \\ & \times
  \, z \, \sqrt{2 \, \partial_+ \, \partial_-} \, K_1 (z \, \sqrt{2 \,
    \partial_+ \, \partial_-}) \, \delta (x^+ - y^+) \, \delta (x^- -
  y^-).
\end{align}
Integrating by parts in \eq{Pij2} we can replace all $\partial_+$ with
\begin{align}
  \partial_+ \, \rightarrow \, - i \, q^-
\end{align}
after which the integral over $x^+ - y^+$ simply cancels $\delta (x^+
- y^+)$. For all factors of $\partial_-$ to the left of the sum in
\eq{Pij2} integration by parts also gives $\partial_- \rightarrow - i
\, q^+$. Finally, the factors of $\partial_-$ to the right of the sum
in \eq{Pij2} are also replaced by $\partial_- \rightarrow - i \, q^+$,
which can be seen by integrating over $y^-$ in \eq{Pij2}, applying
$\partial_-$ which appear to the right of the sum, and undoing the
$y^-$-integral. Finally remembering that $Q^2 = q^2 = - 2 \, q^+ \,
q^- >0$ in DIS we write
\begin{align}\label{Pij3}
  \Pi^{ij} (q^+, & \, q^-, {\un q}=0) \, = \, \frac{\Lambda^2 \,
    S_\perp}{a} \, \frac{N_c^2}{8 \, \pi^2} \, \frac{\mu}{2 \, a} \,
  \delta^{ij} \, Q^2 \, (q^-)^2 \, \int\limits_0^a d x^- \,
  \int\limits_{-\infty}^\infty \, d y^- \, e^{i \, q^+ \,(x^- - y^-)}
  \, \int\limits_0^\infty \, d z \, z^4 \, K_1 (Q \, z ) \notag \\ &
  \times \, \sum\limits_{n=1}^\infty \, \left[ \frac{\mu}{a} \,
    (q^-)^2 \, {\hat G}_{z} (-i \, q^- \, \partial_-) \, z^4 \, \theta
    (x^-) \, \theta (a-x^-) \right]^{n-1} \, z \, \, K_1 (Q \, z) \,
  \delta (x^- - y^-).
\end{align}
We have purposefully did not carry out the $y^-$ integration as the
expression in the form shown in \eq{Pij3} will be easier to evaluate.


\subsubsection{Longitudinal Components of the Hadronic Tensor}

To find the structure functions $\Pi_1$ and $\Pi_2$ from \eq{strf_def}
we need to find one of the longitudinal components ($\Pi^{++}$,
$\Pi^{--}$, or $\Pi^{+-}$) of the hadronic tensor $\Pi^{\mu\nu}$.
Finding only one of the longitudinal components of $\Pi^{\mu\nu}$ is
sufficient, along with \eq{Pij3}, to uniquely determine $\Pi_1$ and
$\Pi_2$. We will determine $\Pi^{++}$.

We start by solving \eq{M-} for $A_-$:
\begin{align}\label{A-}
  A_- (x^+, x^-, z) \, = \, \frac{1}{\partial_+^2} \, \left[
    \partial_+ \, \partial_- + \frac{1}{z} \, \partial_z -
    \partial_z^2 \right] \, A_+ (x^+, x^-, z).
\end{align}
Plugging this into \eq{Mz} we write
\begin{align}\label{A+}
  \left[ 2 \, \partial_+ \, \partial_- - \frac{1}{z^2} + \frac{1}{z}
    \, \partial_z - \partial_z^2 \right] \, \partial_z \, A_+ (x^+,
  x^-, z) \, = \, - \frac{\mu}{a} \, z^4 \, \theta (x^-) \, \theta
  (a-x^-) \, \partial_+^2 \, \partial_z \, A_+ (x^+, x^-, z).
\end{align}

Similar to the transverse field components case above, we begin by
solving Eqs.~(\ref{A-}) and (\ref{A+}) in the $\mu =0$ case of no
shock wave. Solving \eq{A+} for $\mu =0$ while requiring that
$A_+^{(0)} (x^+, x^-,z)$ remains finite as $z \rightarrow +\infty$ we
get
\begin{align}\label{A+0}
  A_+^{(0)} (x^+, x^-, z) \, = \, A_+^b (x^+, x^-) + \frac{1}{2 \,
    \partial_+ \, \partial_-} \, \left[ 1 - z \, \sqrt{2 \, \partial_+
      \, \partial_-} \, K_1 \left( z \, \sqrt{2 \, \partial_+ \,
        \partial_-} \right) \right] \, C (x^+, x^-)
\end{align}
with $C (x^+, x^-)$ an arbitrary function of $x^+$ and $x^-$ and
$A_+^b (x^+, x^-)$ the boundary value of the field $A_+^{(0)} (x^+,
x^-,z)$.  To fix $C (x^+, x^-)$ we plug \eq{A+0} into \eq{A-} and
match terms at $z=0$ obtaining
\begin{align}
  C (x^+, x^-) \, = \, \partial_+^2 \, A_-^b (x^+, x^-) - \partial_+
  \, \partial_- \, A_+^b (x^+, x^-),
\end{align}
where $A_-^b (x^+, x^-)$ is the boundary value of the field $A_-^{(0)}
(x^+, x^-,z)$.  We thus have
\begin{subequations}\label{A+-0}
  \begin{align}
    A_+^{(0)} (x^+, x^-, z) \, = \, \frac{1}{2} \, \left[ 1 + z \,
      \sqrt{2 \, \partial_+ \, \partial_-} \, K_1 \left( z \, \sqrt{2
          \, \partial_+ \, \partial_-} \right) \right] \, A_+^b (x^+,
    x^-) \notag \\ + \frac{\partial_+}{\partial_-} \, \frac{1}{2} \,
    \left[ 1 - z \, \sqrt{2 \, \partial_+ \, \partial_-} \, K_1 \left(
        z \, \sqrt{2 \, \partial_+ \, \partial_-} \right) \right] \,
    A_-^b (x^+, x^-) \\
    A_-^{(0)} (x^+, x^-, z) \, = \, \frac{\partial_-}{\partial_+} \,
    \frac{1}{2} \, \left[ 1 - z \, \sqrt{2 \, \partial_+ \,
        \partial_-} \, K_1 \left( z \, \sqrt{2 \, \partial_+ \,
          \partial_-} \right) \right] \, A_+^b (x^+, x^-) \notag \\ +
    \frac{1}{2} \, \left[ 1 + z \, \sqrt{2 \, \partial_+ \,
        \partial_-} \, K_1 \left( z \, \sqrt{2 \, \partial_+ \,
          \partial_-} \right) \right] \, A_-^b (x^+, x^-).
  \end{align}
\end{subequations}

Using the $\mu$-expansion technique developed above we can write the
solution of \eq{A+} for $\partial_z \, A_+$ as
\begin{align}\label{A+sol}
  \partial_z \, A_+ (x^+, x^-,z) \, = \, \sum\limits_{n=0}^\infty & \,
  \partial_z \, A_+^{(n)} (x^+, x^-,z) \, = \,
  \sum\limits_{n=0}^\infty \, \left[ - \frac{\mu}{a} \, {\hat G}_z^L
    (\partial_+ \, \partial_-) \, z^4 \, \theta (x^-) \, \theta
    (a-x^-) \, \partial_+^2 \right]^n \notag \\
  & \times \, z \, K_0 (z \, \sqrt{2 \, \partial_+ \, \partial_-}) \,
  \left[ \partial_+^2 \, A_-^b (x^+, x^-) - \partial_+ \, \partial_-
    \, A_+^b (x^+, x^-) \right]
\end{align}
where we have defined
\begin{align}\label{GLabr}
  {\hat G}_z^L (\partial_+ \, \partial_-) \, f(x^+, x^-, z) \, = \,
  \int_0^\infty \, \frac{d z'}{z'} \, G_L (z; z'; \partial_+ \,
  \partial_-) \, f(x^+, x^-, z')
\end{align}
with the longitudinal Green function defined by
\begin{align}
   \left[ 2 \, \partial_+ \, \partial_- - \frac{1}{z^2} + \frac{1}{z}
    \, \partial_z - \partial_z^2 \right] \, G_L (z; z'; \partial_+ \,
  \partial_-) \, = \, z' \, \delta (z-z'). 
\end{align}
Requiring that $G_L (z; z'; \partial_+ \, \partial_-)$ goes to zero as
$z \rightarrow 0$ and that it is finite at $z \rightarrow +\infty$ one
readily obtains \cite{Polchinski:2002jw,Hatta:2007cs,Avsar:2009xf}
\begin{align}\label{GLIK}
  G_L (z; z'; \partial_+ \, \partial_-) \, = \, z \, z' \, I_0 \left(
    z_< \, \sqrt{2 \, \partial_+ \, \partial_-} \right) \, K_0 \left(
    z_> \, \sqrt{2 \, \partial_+ \, \partial_-} \right).
\end{align}
The subscript $L$ for Green function $G_L$ stands for longitudinal
components. (Here we are using the same notation as in
\cite{Avsar:2009xf}.) In arriving at \eq{A+sol} we have again demanded
that $A_+^{(n)} (x^+, x^-,z=0) =0$ for $n \ge 1$, such that $A_+ (x^+,
x^-,z=0) = A_+^b (x^+, x^-)$. \eq{A-} can be used together with
\eq{A+sol} to find $A_- (x^+, x^-, z)$ as a series in powers of $\mu$:
one can easily show that $A_-^{(n)} (x^+, x^-,z=0) =0$ for $n \ge 1$
as well.

To evaluate $\Pi^{++}$ we use Eqs. (\ref{Mcl}) and (\ref{Pexpr}) along
with the fact that $A_+^{(n)} (x^+, x^-,z=0) =0$ and $A_-^{(n)} (x^+,
x^-,z=0) =0$ for $n \ge 1$ obtaining
\begin{align}\label{P++1}
  \Pi^{++} (q) \, & = \, - \frac{\Lambda^2}{a} \, \frac{N_c^2}{16 \,
    \pi^2} \int d^2 x_\perp \, d^2 y_\perp \, d x^- \, d y^- \, d (x^+
  - y^+) \, e^{- i \, q \, \cdot \, (x-y)} \int d^4 {\tilde x} \,
  \left[ \frac{1}{z} \frac{\delta A_+^{(0)} ({\tilde x},z )}{\delta
      A_+^b (x) } \, \partial_z \frac{\delta A_- ({\tilde
        x},z )}{\delta A_+^b (y) } \right. \notag \\
  & \left. + \frac{1}{z} \frac{\delta A_-^{(0)} ({\tilde x},z
      )}{\delta A_+^b (x) } \, \partial_z \frac{\delta A_+ ({\tilde
        x},z )}{\delta A_+^b (y) } - \frac{1}{z} \frac{\delta
      A_+^{(0)} ({\tilde x},z )}{\delta A_+^b (x) } \, \partial_z
    \frac{\delta A_-^{(0)} ({\tilde x},z )}{\delta A_+^b (y) } -
    \frac{1}{z} \frac{\delta A_-^{(0)} ({\tilde x},z )}{\delta A_+^b
      (x) } \, \partial_z \frac{\delta A_+^{(0)} ({\tilde x},z
      )}{\delta A_+^b (y) } \right]\Bigg|_{z=0}.
\end{align}
With the help of Eqs.~(\ref{A+-0}), (\ref{A+sol}), (\ref{A+}) and
(\ref{A-}) we write
\begin{subequations}\label{Aders}
\begin{align}
  \frac{\delta A_+^{(0)} (x^+, x^-, z )}{\delta A_+^b (y^+, y^-) } \,
  & = \, \delta (x^+ - y^+) \, \delta (x^- - y^-) + o\left(z^2 \, \ln
    z\right) \label{A+0der} \\
  \frac{\delta A_-^{(0)} (x^+, x^-, z )}{\delta A_+^b (y^+, y^-) } \,
  & = \, o\left( z^2 \, \ln z \right) \label{A-0der} \\
  \frac{\delta A_-^{(n)}(x^+, x^-, z )}{\delta A_+^b (y^+, y^-) } \, &
  = \, - \frac{\partial_-}{\partial_+} \, \frac{\delta A_+^{(n)}(x^+,
    x^-, z )}{\delta A_+^b (y^+, y^-) } + o (z^4) \notag \\ & \, = \,
  - z^2 \, \frac{\mu}{2 \, a} \, \frac{\partial_-}{\partial_+} \,
  \int\limits_0^\infty \, d z' \, K_0 (z' \, \sqrt{2 \, \partial_+ \,
    \partial_-}) \, z'^4 \,
  \theta (x^-) \, \theta (a-x^-) \, \partial_+^2 \notag \\
  & \times \, \left[ - \frac{\mu}{a} \, {\hat G}_{z'}^L (\partial_+ \,
    \partial_-) \, z'^4 \, \theta (x^-) \, \theta (a-x^-) \,
    \partial_+^2 \right]^{n-1} \, z' \, K_0 (z' \, \sqrt{2 \,
    \partial_+ \, \partial_-}) \, \partial_+ \, \partial_- \notag \\ &
  \times \, \delta (x^+ - y^+) \, \delta (x^- - y^-) + o\left(z^4
  \right) \label{A+nder}
\end{align}
\end{subequations}
with \eq{A+nder} valid for $n \ge 1$.

Using Eqs.~(\ref{Aders}) in \eq{P++1} and integrating over $\tilde x$,
$x_\perp$, $y_\perp$, and $x^+ - y^+$ similar to how it was done in
arriving at \eq{Pij3} yields
\begin{align}\label{P++2}
  \Pi^{++} (q^+, & \, q^-, {\un q}=0) \, = \, \frac{\Lambda^2 \,
    S_\perp}{a} \, \frac{N_c^2}{32 \, \pi^2} \, \frac{\mu}{2 \, a} \,
  Q^4 \, \int\limits_0^a d x^- \, \int\limits_{-\infty}^\infty d y^-
  \, e^{i \, q^+ \, (x^- - y^-)} \, \int\limits_0^\infty \, d z
  \, z^4 \, K_0 (z \, Q) \notag \\
  & \times \, \sum\limits_{n=1}^\infty \, \left[ \frac{\mu}{a} \,
    (q^-)^2 \, {\hat G}_{z}^L (- i \, q^- \, \partial_-) \, z^4 \,
    \theta (x^-) \, \theta (a-x^-) \right]^{n-1} \, z \, K_0 (z \, Q)
  \, \delta (x^- - y^-).
\end{align}

Eqs. (\ref{Pij3}) and (\ref{P++2}) give us the most general
expressions for the two components of the hadronic tensor that we need
to determine the structure functions $\Pi_1$ and $\Pi_2$. We will now
evaluate them in the eikonal approximation.


\subsection{Eikonal Approximation and Its Applicability Region}
\label{sec:eik}


\subsubsection{Eikonal Hadronic Tensor}

Let us evaluate the expressions (\ref{Pij3}) and (\ref{P++2}) in the
eikonal approximation first. Eikonal approximation corresponds to the
mathematical limit of $a \rightarrow 0$, such that \eq{Phi} becomes
\begin{align}\label{eik_lim}
  \Phi (x^-) \, \rightarrow \, \mu \, \delta (x^-). 
\end{align}
Eikonal approximation in AdS/CFT has been employed before in
\cite{Brower:2007qh,Cornalba:2007zb,Albacete:2009ji,Levin:2008vj,Mueller:2008bt,Hatta:2007he,Hatta:2007cs}.
The eikonal approximation (\ref{eik_lim}) tries to mimic the physical
limit when the shock wave is infinitely boosted. Indeed for a
real-life proton or nucleus the physical limit of infinite boost can
be achieved by sending the large momentum of the proton/nucleus wave
$p^+$ to infinity. While indeed in the $p^+ \rightarrow \infty$ limit
$a \rightarrow 0$ as follows from \eq{a}, one also notices from
\eq{mu} that strictly-speaking in this limit $\mu \rightarrow \infty$
making \eq{eik_lim} meaningless. We will understand the eikonal limit
as the case when $p^+$ is very large but is still finite, such that
the delta-function approximation of \eq{eik_lim} is valid, though
$\mu$ is very large. The eikonal approximation can be thought of as
taking $p^+ \rightarrow \infty$ limit, while simultaneously sending
$\Lambda \rightarrow 0$ in such a way that $\mu$ would remain
constant, the possibility of which follows from \eq{mu}. This is
similar to Aichelburg and Sexl's construction of ultrarelativistic
black hole metric \cite{Aichelburg:1970dh}.  Mathematically the
eikonal limit is simply equivalent to taking $a \rightarrow 0$ while
keeping $\mu$ fixed.
\FIGURE{\includegraphics[width=15cm]{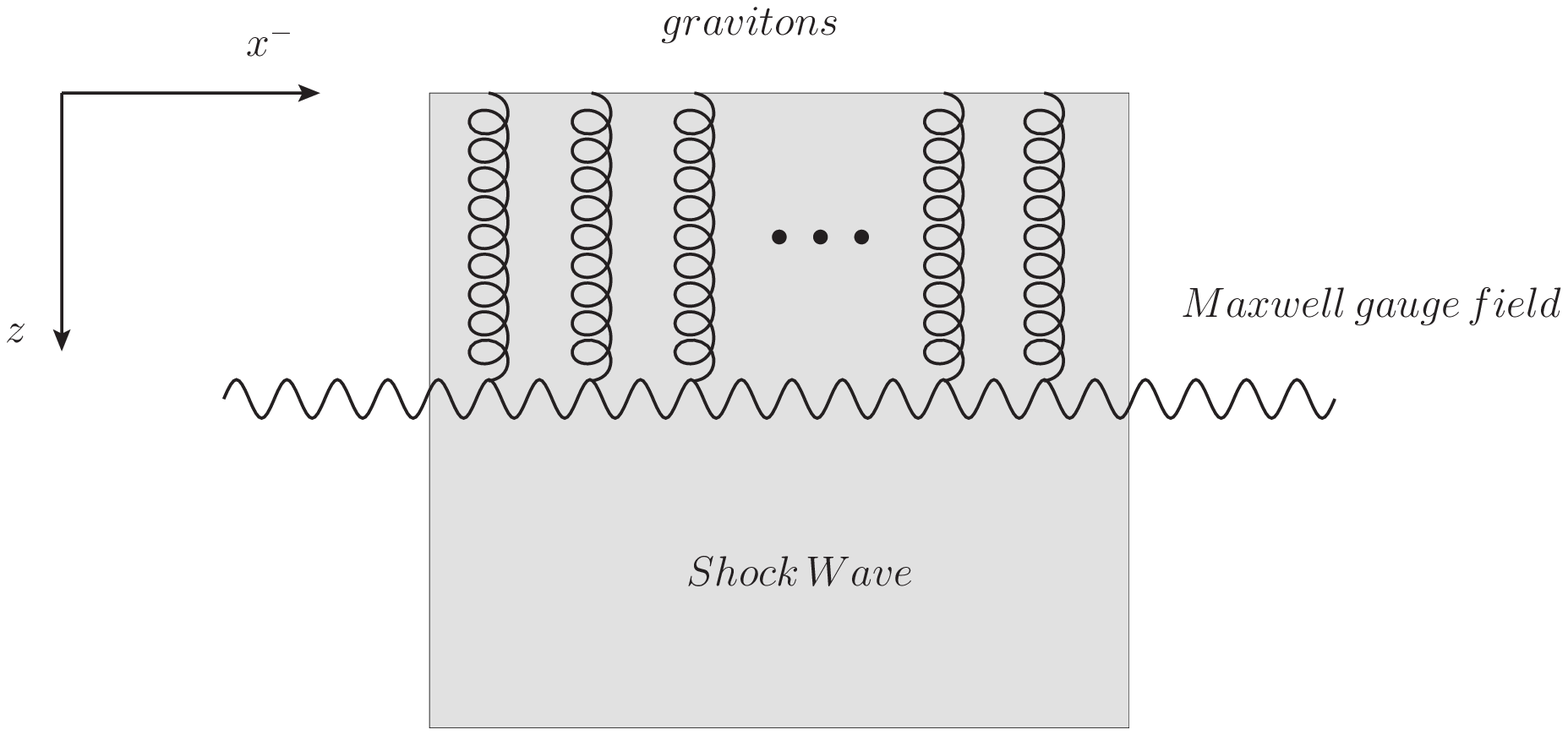}
  \caption{Diagrammatic representation of the Maxwell gauge field scattering 
    on the shock wave in the bulk.  The horizontal wiggly line
    represents the gauge field, while the vertical cork-screw lines
    represent graviton exchanges with the shock wave. The boundary of
    AdS$_5$ is at the top of the shock wave, which in turn is denoted
    by the shaded rectangle. }
  \label{mult}
}

Following \cite{Albacete:2008vs,Albacete:2009ji} one can identify the
series in Eqs. (\ref{Pij3}) and (\ref{P++2}) with the scattering of
the gauge field in the graviton field of a shock wave. Each power of
$\mu$ in Eqs. (\ref{Pij3}) and (\ref{P++2}) corresponds to a graviton
exchange with the shock wave.\footnote{An interesting (but irrelevant
  for presented calculations) question is where exactly the graviton
  field of the shock wave originates: since our shock wave (\ref{nuc})
  has no source in the bulk, one can think of is as having a source at
  $z = \infty$ (see \cite{Kovchegov:2009du}) with graviton exchanges
  with that source, or one may think of the boundary condition at
  $z=0$ (that we have a nucleus in four dimensions) as being the
  effective ``source'' for the shock wave, with the gravitons
  exchanged with the boundary, as shown in \fig{mult}.} The scattering
of the gauge field in the shock wave is shown in \fig{mult}. If $a$ is
small the shock wave has a very short extent in the $x^-$-direction,
such that the derivative $\partial^-$ in \eq{Green} is very large. We
thus approximate the eikonal Green function as (see
\cite{Albacete:2009ji} for a similar approach to shock wave scattering
with the goal of modeling heavy ion collisions)
\begin{align}
  G^{eik} (z; z'; \partial_+ \, \partial_-) \, = \, \frac{1}{2 \,
    \partial_+ \, \partial_-} \, z' \, \delta (z-z').
\end{align}
One then has
\begin{align}\label{Gabr_eik}
  {\hat G}_z^{eik} (\partial_+ \, \partial_-) \, f(x^+, x^-, z) \, =
  \, \frac{1}{2 \, \partial_+ \, \partial_-} \, f(x^+, x^-, z).
\end{align}

We start by evaluating the transverse components of the hadronic
tensor first.  Replacing $\partial_+ \rightarrow - i \, q^-$ in
\eq{Gabr_eik} and using the result in \eq{Pij3} yields
\begin{align}\label{Pij1_eik}
  \Pi^{ij}_{eik} (q^+, & \, q^-, {\un q}=0) \, = \, \frac{\Lambda^2 \,
    S_\perp}{a} \, \frac{N_c^2}{8 \, \pi^2} \, \frac{\mu}{2 \, a} \,
  \delta^{ij} \, Q^2 \, (q^-)^2 \, \int\limits_0^a d x^- \,
  \int\limits_{-\infty}^\infty \, d y^- \, e^{i \, q^+ \,(x^- - y^-)}
  \, \int\limits_0^\infty \, d z \, z^4 \, K_1 (Q \, z ) \notag \\ &
  \times \, \sum\limits_{n=1}^\infty \, \left[ i \, \frac{\mu}{2 \, a}
    \, q^- \, \frac{1}{\partial_-} \, z^4 \, \theta (x^-) \, \theta
    (a-x^-) \right]^{n-1} \, z \, \, K_1 (Q \, z) \, \delta (x^- -
  y^-).
\end{align}

Now since for $n \ge 2$ and for $x^- \le a$
\begin{align}
  \left[ \frac{1}{\partial_-} \, \theta (x^-) \, \theta (a-x^-)
  \right]^{n-1} \, \delta (x^- - y^-) \, = \, \theta (y^-) \, \theta
  (a-y^-) \, \theta (x^- - y^-) \, \frac{(x^- - y^-)^{n-2}}{(n-2)!}
\end{align}
we obtain
\begin{align}\label{eik_long}
  \int\limits_0^a d x^- \, \int\limits_{-\infty}^\infty & \, d y^- \,
  e^{i \, q^+ \,(x^- - y^-)} \, \left[ \frac{1}{\partial_-} \, \theta
    (x^-) \, \theta (a-x^-) \right]^{n-1} \, \delta (x^- - y^-) \notag
  \\ \, & = \, \frac{a^n}{n!} \, \sum\limits_{m=0}^\infty \, \frac{(i
    \, q^+ \, a)^m}{m!} \, \frac{(n-1) \, n}{(n+m-1) \, (n+m)} \notag
  \\ \, & = \, \frac{a^n}{(n-1)!} \, \left\{ e^{i \, q^+ \, a} - (- i
    \, q^+ \, a)^{-n} \, (i \, q^+ \, a + n -1) \, \left[ \Gamma (n) -
      \Gamma (n, - i \, q^+ \, a) \right] \right\}.
\end{align}
The $n$th term in the series of \eq{Pij1_eik} brings in a factor of
$\mu^n/a^{n+1}$ on top of the factor we obtained in \eq{eik_long}.
Note that one factor of $1/a$ comes from the prefactor in the
definition of $\Pi^{\mu\nu}$ in \eq{Pdef}. Therefore, the $a
\rightarrow 0$ eikonal limit should not apply to this factor.  For the
purpose of the eikonal approximation the $n$th term in the series of
\eq{Pij1_eik} is then of the order $\mu^n/a^{n}$ times \eq{eik_long}.
It is then clear that in the $a \rightarrow 0$ limit with $\mu$ fixed
only the first term in the series in the second line of \eq{eik_long}
survives, as
\begin{align}\label{eik_long1}
  \int\limits_0^a d x^- \, \int\limits_{-\infty}^\infty & \, d y^- \,
  e^{i \, q^+ \,(x^- - y^-)} \, \left[ \frac{1}{\partial_-} \, \theta
    (x^-) \, \theta (a-x^-) \right]^{n-1} \, \delta (x^- - y^-) \, =
  \, \frac{a^n}{n!} + o \left( a^{n+1} \right).
\end{align}
Using the eikonal approximation of \eq{eik_long1} in \eq{Pij1_eik} we
obtain
\begin{align}\label{Pij2_eik}
  \Pi^{ij}_{eik} (q^+, q^-, {\un q}=0) \, = \, i \, \frac{\Lambda^2 \,
    S_\perp}{a} \, \frac{N_c^2}{8 \, \pi^2} \, \delta^{ij} \, Q^2 \,
  q^- \int\limits_0^\infty \, d z \, z \, \left[ K_1 (Q \, z )
  \right]^2 \, \left[ 1 - \exp \left( \frac{i}{2} \, \mu \, q^- \, z^4
    \right) \right]
\end{align}
in agreement with the eikonal formulas used in
\cite{Mueller:2008bt,Levin:2008vj} and recently derived in
\cite{Avsar:2009xf} for DIS on a shock wave. Remembering that $a =
A^{1/3} /p^+$ we see that $q^-/a = p^+ q^- /A^{1/3} = Q^2 / (2 \, x \,
A^{1/3})$ such that \eq{Pij2_eik} can be re-written as
\begin{align}\label{Pij3_eik}
  \Pi^{ij}_{eik} (q^+, q^-, {\un q}=0) \, = \, i \, \Lambda^2 \,
  S_\perp \, \frac{N_c^2}{16 \, \pi^2} \, \delta^{ij} \, \frac{Q^4}{x
    \, A^{1/3}} \int\limits_0^\infty \, d z \, z \, \left[ K_1 (Q \, z
    ) \right]^2 \, \left[ 1 - \exp \left( \frac{i}{2} \, \mu \, q^- \,
      z^4 \right) \right]
\end{align}
making the agreement with \cite{Avsar:2009xf} manifest (up to a
trivial factor of $\Lambda^2 \, S_\perp$ which probably signifies a
slightly different overall normalization used in \cite{Avsar:2009xf}).

Similar calculations for $\Pi^{++}$ from \eq{P++2} give the eikonal
expression
\begin{align}\label{P++eik}
  \Pi^{++}_{eik} (q^+, q^-, {\un q}=0) \, = \, i \, \Lambda^2 \,
  S_\perp \, \frac{N_c^2}{16 \, \pi^2} \, \frac{Q^2}{4 \, (q^-)^2} \,
  \frac{Q^4}{x \, A^{1/3}} \int\limits_0^\infty \, d z \, z \, \left[
    K_0 (Q \, z ) \right]^2 \, \left[ 1 - \exp \left( \frac{i}{2} \,
      \mu \, q^- \, z^4 \right) \right].
\end{align}
\eq{P++eik} agrees with the result of \cite{Avsar:2009xf} up to the
same overall normalization factor of $\Lambda^2 \, S_\perp$ as for
$\Pi^{ij}_{eik}$.

Note that the eikonal approximation developed in
\cite{Albacete:2009ji} also leads to the eikonal propagator (the
factor in the square brackets) in Eqs.~(\ref{Pij3_eik}) and
(\ref{P++eik}) originally obtained in
\cite{Mueller:2008bt,Levin:2008vj}. To see this note that the
truncated eikonal graviton amplitude in Eq. (3.29) of
\cite{Albacete:2009ji} is proportional to
\begin{align}\label{asym}
  \sum\limits_{n=0}^\infty \, (n+1) \, \left( - \frac{1}{2} \, z^4 \,
    t_2 (x^+) \, \frac{\partial_-}{\partial_+} \right)^n \, t_2 (x^+),
\end{align}
where, for the purposes of comparing with \eq{Pij3_eik} we take the
shock wave profile to be
\begin{align}\label{t2}
  t_2 (x^+) \, = \, \frac{\mu_2}{a_2} \, \theta (x^+) \, \theta (a-x^+).
\end{align}
In \cite{Albacete:2009ji} a proton-nucleus collision was modeled by
colliding a shock wave with a smaller energy density (a proton) with a
shock wave with a larger energy density (a nucleus). The larger
``nucleus'' shock wave was chosen to move in the $x^-$ direction in
\cite{Albacete:2009ji}: this is why its profile in \eq{t2} is a
function of $x^+$ instead of $x^-$.

Using \eq{t2} one can readily show that
\begin{align}
  \lim_{a_2 \rightarrow 0} \left( t_2 (x^+) \, \frac{1}{\partial_+}
  \right)^n \, t_2 (x^+) \, = \, \frac{\mu_2^{n+1}}{(n+1)!} \, \delta
  (x^+),
\end{align}
which, when substituted into \eq{asym} after summing the series over
$n$ yields
\begin{align}\label{hbarexp}
  \exp \left( - \frac{1}{2} \, z^4 \, \mu_2 \, \partial_- \right) \,
  t_2 (x^+).
\end{align}
Identifying $\partial_-$ in \eq{hbarexp} with $- i \, q^+$ in
Eqs.~(\ref{Pij3_eik}) and (\ref{P++eik}) we see that the exponents in
both equations are identical up to $x^+ \leftrightarrow x^-$
interchange. (In \cite{Albacete:2009ji} the series started from $n=0$
since $n=0$-term corresponded to no additional rescatterings, but
still contained graviton production, in contrast to the DIS case at
hand, where no rescattering implies no interactions and hence no
contribution to DIS cross section.  This is why we do not subtract $1$
from the exponent in \eq{hbarexp} unlike the exponents in
Eqs.~(\ref{Pij3_eik}) and (\ref{P++eik}).)

Finally, as $\mu = p^+ \, \Lambda^2 \, A^{1/3}$ we see that $\mu \,
q^- = Q^2 \, \Lambda^2 \, A^{1/3}/(2 \, x)$. Defining a momentum
scale\footnote{As one can see from Eqs.~(\ref{mu}) and (\ref{a}) the
  expressions we use for $\mu$ and $a$ are accurate up to a constant
  and possibly some factors of 't Hooft coupling $\lambda$
  \cite{Albacete:2008vs}. Such factors, which are important for
  phenomenology, can be easily included later.}
\begin{align}\label{Q1}
  Q_1^2 (x, A) \, \equiv \, \frac{\Lambda^2 \, A^{1/3}}{4 \, x}
\end{align}
we recast Eqs.~(\ref{Pij3_eik}) and (\ref{P++eik}) into the following
form:
\begin{subequations}\label{Pij4_eik}
\begin{align}
  \Pi^{ij}_{eik} (q^+, q^-, {\un q}=0) \, & = \, i \, \Lambda^2 \,
  S_\perp \, \frac{N_c^2}{16 \, \pi^2} \, \delta^{ij} \, \frac{Q^4}{x
    \, A^{1/3}} \int\limits_0^\infty \, d z \, z \, \left[ K_1 (Q \, z
    ) \right]^2 \, \left[ 1 - \exp \left( i \, Q^2 \, Q_1^2 (x, A) \,
      z^4 \right) \right], \\
  \Pi^{++}_{eik} (q^+, q^-, {\un q}=0) \, & = \, i \, \Lambda^2 \,
  S_\perp \, \frac{N_c^2}{16 \, \pi^2} \, \frac{Q^2}{4 \, (q^-)^2} \,
  \frac{Q^4}{x \, A^{1/3}} \int\limits_0^\infty \, d z \, z \, \left[
    K_0 (Q \, z ) \right]^2 \, \left[ 1 - \exp \left( i \, Q^2 \,
      Q_1^2 (x, A) \, z^4 \right) \right].
\end{align}
\end{subequations}
In \cite{Mueller:2008bt,Levin:2008vj,Avsar:2009xf} the scale $Q_1^2$
was identified with the saturation scale $Q_s^2$ as in the eikonal
approximation it is the only momentum scale in the problem and because
structure functions become independent of $x$ for $Q^2 < Q_1^2$, in
agreement with what one observes in perturbative approaches
\cite{Kovchegov:1999ua}. Our understanding of the physical meaning of
$Q_1^2$ will be detailed in Sec. \ref{MS}.


\subsubsection{Applicability Region of the Eikonal Approximation}
\label{appl}

The question we would like to address now is whether the eikonal
result (\ref{Pij4_eik}) gives us the complete ``hadronic tensor''
$\Pi^{\mu\nu}$ in the small-$x$ limit. As we noted above, the proper
physical high energy limit corresponds to increasing the
proton/nucleon momentum $p^+$, and not to simply taking $a \rightarrow
0$ limit. In arriving at Eqs.~(\ref{Pij4_eik}) we have made several
approximations. In particular we have neglected higher powers of $q^+
\, a$ in approximating \eq{eik_long} with \eq{eik_long1}. Since
\begin{align}\label{qa}
  q^+ \, a \, = \, \frac{q^+}{p^+} \, A^{1/3} \, = \, - x \, A^{1/3}
\end{align}
we are neglecting higher powers of Bjorken-$x$, which seems to be
justified in the small-$x$ limit.\footnote{Note that $q^+ \, a \sim a
  / l_{coh}$, where $l_{coh}$ is the coherence length of the
  projectile (particles the $R$-current decays into) in the
  $x^-$-direction: smallness of $q^+ \, a$ means that the coherence
  length is much larger than the size of the proton/nucleus.} However,
it is easy to see from Eqs.~(\ref{Pij4_eik}) that if we expand the
exponentials in them back into series, the series would be in the
powers of
\begin{align}
  \frac{Q_1^2 (x, A)}{Q^2} \, \sim \, \frac{\Lambda^2 \, A^{1/3}}{Q^2
    \, x}. 
\end{align}
We would then have
\begin{align}\label{Power}
  \Pi^{\mu\nu} (q^+, q^-, {\un q}=0) \, \sim \, \sum\limits_n \, c_n
  \, \left[ \frac{\Lambda^2 \, A^{1/3}}{Q^2 \, x} \right]^n \, \left[
    1 + d_n^1 \, x \, A^{1/3} + d_n^2 \, (x \, A^{1/3})^2 + \ldots
  \right]
\end{align}
with $c_n$'s and $d_n^m$'s some $x$- and $Q^2$-independent constants.
We now have a problem: if we want to take a small-$x$/fixed-$Q^2$
limit (which is the same as taking $p^+$ large in Eqs.~(\ref{mu}) and
(\ref{a})), the eikonal formulas (\ref{Pij4_eik}) would receive
order-1 corrections. Namely, if we do power counting in $x$ in
\eq{Power}, we see that subleading order-$x$ non-eikonal correction to
the $(n+1)$st term is of the same order as the $n$th term in the
eikonal series
\begin{align}\label{Power1}
  c_{n+1} \, \left[ \frac{\Lambda^2 \, A^{1/3}}{Q^2 \, x}
  \right]^{n+1} \, d_{n+1}^1 \, x \, A^{1/3} \, \sim \, c_n \, \left[
    \frac{\Lambda^2 \, A^{1/3}}{Q^2 \, x} \right]^n
\end{align}
as each of them is of the order $x^{-n}$. The parametric equality can
be simplified to
\begin{align}
  \frac{\Lambda^2 \, A^{2/3}}{Q^2} \sim 1. 
\end{align}
We see that the non-eikonal corrections become important at $Q \sim
Q_2$ with the momentum scale $Q_2$ defined by
\begin{align}\label{Q2}
  Q_2 (A) \, \equiv \, \Lambda \, A^{1/3}.
\end{align}
This scale is similar to the saturation scale identified in scattering
a dipole on a shock wave in \cite{Albacete:2008ze} (see Eq. (4.28)
there) and also in \cite{Dominguez:2008vd}. The only difference
between $Q_2$ and the saturation scale found in \cite{Albacete:2008ze}
is a factor of $\sqrt{\lambda}$ contained in the latter which appears
not to be present in $Q_2$ ($\lambda$ is the 't~Hooft coupling
constant).  This factor of $\sqrt{\lambda}$ is inherently present in
any calculation of the dynamics of a fundamental superstring as it is
a prefactor in the Nambu-Goto action (see e.g.
\cite{Maldacena:1998im}).  At the same time any AdS/CFT-based
calculation of $R$-current correlators in the large-$\lambda$ limit
appears to be $\lambda$-independent. To date we have not found a
satisfactory explanation of this difference by $\sqrt{\lambda}$ and
suspect that it may be related to some more fundamental questions
concerning AdS/CFT correspondence. We leave this question open for
further research.\footnote{Indeed $\mu$ may depend on $\lambda$, as
  was suggested in Appendix A of \cite{Albacete:2008vs}: however this
  would not explain the difference between \eq{Q2} and Eq. (4.28) in
  \cite{Albacete:2008ze}, as $\lambda$-dependence in $\mu$ would
  modify both of them in the same way.}

A priori the scale $Q_2$ from \eq{Q2} is not the only scale one can
construct out of non-eikonal corrections in \eq{Power}. Equating the
$k$th non-eikonal correction to the $(n+m)$th term in the series of
\eq{Power} to the $n$th eikonal term yields\footnote{It may happen
  that one of these terms is real, while the other one is purely
  imaginary: if one then insists on equating only real terms to real
  terms and imaginary terms to imaginary terms, one should equate the
  $n$th eikonal term to the $(2 \, k)$th correction in the $(n + 2 \,
  m)$th term, obtaining the same result as below. Powers of $i$ in
  Eqs.  (\ref{eik_long}) and (\ref{Pij4_eik}) would insure that the
  terms we compare are either both real or both imaginary. }
\begin{align}\label{Power2}
  c_{n+m} \, \left[ \frac{\Lambda^2 \, A^{1/3}}{Q^2 \, x}
  \right]^{n+m} \, d_{n+m}^k \, \left( x \, A^{1/3} \right)^k \, \sim
  \, c_n \, \left[ \frac{\Lambda^2 \, A^{1/3}}{Q^2 \, x} \right]^n
\end{align}
giving a possible new scale at which non-eikonal corrections should
become important:
\begin{align}\label{Qmk}
  Q_{m,k}^2 \, \sim \, \frac{\Lambda^2 \, A^{1/3}}{x} \, \left( x \,
    A^{1/3} \right)^{k/m}.
\end{align}
Adjusting positive integers $k$ and $m$ in \eq{Qmk} one can get the
scale $Q_{m,k}$ as (parametrically) close to $Q_1$ as desirable.
Still, as $k,m \ge 0$ we have $Q_{m,k} \le Q_1$. In fact $Q_{m,k}$'s
are a multitude of scales below $Q_1$ and both above and below $Q_2$.
Since $Q_{m,k}$'s are the scales at which non-eikonal corrections are
important, we conclude that, at least with this a priori analysis, one
can not trust the eikonal formulas (\ref{Pij4_eik}) for $Q \sim
Q_{m,k} \le Q_1$.

Therefore, the conclusion of our power-counting analysis is that the
eikonal expressions in Eqs.~(\ref{Pij4_eik}) are valid only at
\begin{align}
  Q^2 \, \gtrsim \, Q_1^2 (x, A)
\end{align}
or, strictly speaking, only for $Q^2 \, \gg \, Q_1^2 (x, A)$. To
clarify whether this really is the region of validity of the eikonal
approximation or whether it may actually be applicable at $Q^2 < Q_1^2
(x, A)$ one has to find the exact expression for the hadronic tensor.

Note that since
\begin{align}
  Q_{m,k} \, \sim \, Q_1^{1 - \frac{k}{m}} \, Q_2^\frac{k}{m}
\end{align}
the problem of R-current DIS on a shock wave remains a problem with
only {\sl two} momentum scales $Q_1$ and $Q_2$. The series
(\ref{Power}) can be written in terms of these two scales as
\begin{align}\label{Power3}
  \Pi^{\mu\nu} (q^+, q^-, {\un q}=0) \, \sim \, \sum\limits_n \, c_n
  \, \left[ \frac{Q_1^2}{Q^2} \right]^n \, \left[ 1 + d_n^1 \, \left(
      \frac{Q_2}{Q_1} \right)^2 + d_n^2 \, \left( \frac{Q_2}{Q_1}
    \right)^4 + \ldots \right].
\end{align}
The important conclusion of this Subsection is that the DIS process at
strong coupling is a {\sl two-scale problem}.


\subsection{Beyond the Eikonal Approximation: Perturbative Solution}
\label{sec:beyond}

Let us construct the hadronic structure tensor perturbatively by
exactly calculating the terms in the series of Eqs.~(\ref{Pij3}) and
(\ref{P++2}) order-by-order in $\mu$. We denote the $n$th term in each
of those series by
\begin{align}\label{n-term}
  \Pi_{ij}^{(n)} (q^+, & \, q^-, {\un q}=0) \, = \, \frac{\Lambda^2 \,
    S_\perp}{a} \, \frac{N_c^2}{8 \, \pi^2} \, \frac{\mu}{2 \, a} \,
  \delta_{ij} \, Q^2 \, (q^-)^2 \, \int\limits_0^a d x^- \,
  \int\limits_{-\infty}^\infty \, d y^- \, e^{i \, q^+ \,(x^- - y^-)}
  \, \int\limits_0^\infty \, d z \, z^4 \, K_1 (Q \, z ) \notag \\ &
  \times \, \left[ \frac{\mu}{a} \, (q^-)^2 \, {\hat G}_{z} (-i \, q^-
    \, \partial_-) \, z^4 \, \theta (x^-) \, \theta (a-x^-)
  \right]^{n-1} \, z \, \, K_1 (Q \, z) \, \delta (x^- - y^-)
\end{align}
and
\begin{align}\label{n-term++}
  \Pi_{--}^{(n)} (q^+, & \, q^-, {\un q}=0) \, = \, \frac{\Lambda^2 \,
    S_\perp}{a} \, \frac{N_c^2}{32 \, \pi^2} \, \frac{\mu}{2 \, a} \,
  Q^4 \, \int\limits_0^a d x^- \, \int\limits_{-\infty}^\infty d y^-
  \, e^{i \, q^+ \, (x^- - y^-)} \, \int\limits_0^\infty \, d z
  \, z^4 \, K_0 (z \, Q) \notag \\
  & \times \, \left[ \frac{\mu}{a} \, (q^-)^2 \, {\hat G}_{z}^L (- i
    \, q^- \, \partial_-) \, z^4 \, \theta (x^-) \, \theta (a-x^-)
  \right]^{n-1} \, z \, K_0 (z \, Q) \, \delta (x^- - y^-)
\end{align}
with $n = 1, 2 , \ldots$. Below we estimate the $n=1$ and $n=2$ terms.


\subsubsection{Leading Order}

For the $n=1$ term a quick calculation readily gives the leading-order
(LO) terms
\begin{align}\label{1-term}
  \Pi_{ij}^{(1)} \, = \, \delta_{ij} \, \Lambda^2 \, S_\perp \,
  \frac{N_c^2}{10 \, \pi^2} \, \frac{\mu}{a} \, \frac{(q^-)^2}{Q^4}, \ 
  \ \ \ \ \Pi_{--}^{(1)} \, = \, \Lambda^2 \, S_\perp \,
  \frac{N_c^2}{60 \, \pi^2} \, \frac{\mu}{a} \, \frac{1}{Q^2}.
\end{align}
The same expressions would be obtained if one expands the eikonal
formulas (\ref{Pij3_eik}) and (\ref{P++eik}) to order-$\mu$. Indeed
the eikonal approximation of the previous Subsection only modifies the
gauge field propagators sandwiched between the graviton exchanges in
\fig{mult} along with the interaction vertices of \fig{mult}, since we
only modify the Green function as shown in \eq{Gabr_eik} and the
longitudinal integrals over positions of graviton-gauge field
vertices, as follows from \eq{eik_long1}. For $n=1$ we only have one
graviton exchange: the propagators to the left and to the right of the
graviton-gauge field vertex (each of which giving $K_1 (Q \, z )$ in
\eq{Pij3} and $K_0 (Q \, z )$ in \eq{P++2}) are exact, even in the
eikonal approximation, and the longitudinal integral (\ref{eik_long})
is carried out exactly in this case.  Therefore the exact one-graviton
exchange results in \eq{1-term} are the same as the order-$\mu$ terms
in the eikonal formulas (\ref{Pij3_eik}) and (\ref{P++eik}).

As we did above we replace $\mu$ and $a$ using
\begin{align}\label{q/a}
  \frac{q^-}{a} \,  =  \, \frac{Q^2}{2 \, x \, A^{1/3}}
\end{align}
and
\begin{align}\label{mq}
  \mu \, q^- \, = \, \frac{Q^2 \, \Lambda^2 \, A^{1/3}}{2 \, x}.
\end{align}
\eq{1-term} can then be written in terms of $x$ and $Q^2$ as 
\begin{align}\label{1-term2}
  \Pi_{ij}^{(1)} \, = \, \delta_{ij} \, \Lambda^2 \, S_\perp \,
  \frac{N_c^2}{40 \, \pi^2} \, \frac{\Lambda^2}{x^2}, \ \ \ \ \ 
  \Pi_{--}^{(1)} \, = \, \Lambda^2 \, S_\perp \, \frac{N_c^2}{240 \,
    \pi^2} \, \frac{Q^2}{(q^-)^2} \, \frac{\Lambda^2}{x^2}.
\end{align}


\subsubsection{Next-to-Leading Order}

We start by analyzing the transverse components of $\Pi^{\mu\nu}$.  To
find the transverse hadronic tensor at the next-to-leading order (NLO)
we put $n=2$ in \eq{n-term} and employ the definition of the Green
function in Eqs. (\ref{Gabr}) and (\ref{GIK}) obtaining
\begin{align}\label{2-term1}
  \Pi_{ij}^{(2)} & \, = \, \delta_{ij} \, \frac{\Lambda^2 \,
    S_\perp}{a} \, \frac{N_c^2}{16 \, \pi^2} \, \left( \frac{\mu}{a}
  \right)^2 \, Q^2 \, (q^-)^4 \, \int\limits_0^a d x^- \,
  \int\limits_0^a \, d y^- \, e^{i \, q^+ \,(x^- - y^-)} \,
  \int\limits_0^\infty \, d z \, z^5 \, K_1 (z \, Q) \notag \\ &
  \times \, \int\limits_0^\infty d z' \, I_1 \left( z_< \, \sqrt{-i \,
      2 \, q^- \, \partial_-} \right) \, K_1 \left( z_> \, \sqrt{-i \,
      2 \, q^- \, \partial_-} \right) \, z'^5 \, \, K_1 (z' \, Q) \,
  \delta (x^- - y^-)
\end{align}
where we have used $\delta (x^- - y^-)$ to replace $\theta (x^-) \,
\theta (a-x^-)$ by $\theta (y^-) \, \theta (a-y^-)$ and employed the
latter to modify the limits of $y^-$-integration. As before
$\partial_- = \partial / \partial x^-$ and $z_{> \, (<)} \, = \,
\text{max} \, (\text{min}) \, \{ z, z'\}$. 

We write
\begin{align}\label{dint}
  \delta (x^- - y^-) \, = \, \int\limits_{-\infty}^\infty \frac{d \,
    l^+}{2 \, \pi} \, e^{- i \, (l^+ + i \, \epsilon) \, (x^- - y^-)}.
\end{align}
The $+ i \, \epsilon$ regulator is inserted to impose causality: it
makes sure that
\begin{align}
  \frac{1}{\partial_-} \, \delta (x^- - y^-) \, = \, \theta (x^- - y^-)
\end{align}
for $1/\partial_-$ defined in \eq{ints}. Using \eq{dint} we rewrite
\eq{2-term1} as
\begin{align}\label{2-term2}
  \Pi_{ij}^{(2)} & \, = \, \delta_{ij} \, \frac{\Lambda^2 \,
    S_\perp}{a} \, \frac{N_c^2}{16 \, \pi^2} \, \left( \frac{\mu}{a}
  \right)^2 \, Q^2 \, (q^-)^4 \, \int\limits_{-\infty}^\infty \frac{d
    \, l^+}{2 \, \pi} \, \int\limits_0^a d x^- \, \int\limits_0^a \, d
  y^- \, e^{i \, (q^+ - l^+ - i \, \epsilon) \,(x^- - y^-)} \,
  \int\limits_0^\infty \, d z \, z^5 \, K_1 (z \, Q) \notag \\ &
  \times \, \int\limits_0^\infty d z' \, I_1 \left( z_< \, \sqrt{- \,
      2 \, q^- \, (l^+ + i \, \epsilon)} \right) \, K_1 \left( z_> \,
    \sqrt{- \, 2 \, q^- \, (l^+ + i \, \epsilon)} \right) \, z'^5 \,
  \, K_1 (z' \, Q).
\end{align}
Performing $x^-$ and $y^-$ integrations in \eq{2-term2} yields
\begin{align}\label{2-term3}
  \Pi_{ij}^{(2)} & \, = \, \delta_{ij} \, \frac{\Lambda^2 \,
    S_\perp}{a} \, \frac{N_c^2}{16 \, \pi^2} \, \left( \frac{\mu}{a}
  \right)^2 \, \frac{Q^2 \, (q^-)^4}{|q^+|} \,
  \int\limits_{-\infty}^\infty \frac{d \, \xi}{2 \, \pi} \,
  \frac{1}{(1-\xi + i \, \epsilon)^2} \, \left[ 2 - e^{- i \, q^+ \, a
      \, (1 - \xi + i \, \epsilon)} - e^{i \, q^+ \, a \, (1 - \xi + i
      \, \epsilon)} \right] \notag \\ & \times \, \int\limits_0^\infty
  \, d z \, z^5 \, K_1 (z \, Q) \, \int\limits_0^\infty d z' \, I_1
  \left( z_< \, Q \, \sqrt{\xi - i \, \epsilon} \right) \, K_1 \left(
    z_> \, Q \, \sqrt{\xi - i \, \epsilon} \right) \, z'^5 \, \, K_1
  (z' \, Q).
\end{align}
We have defined
\begin{align}
  \xi \, \equiv \, \frac{l^+}{q^+}. 
\end{align}
In arriving at \eq{2-term3} we have also used the fact that $0 < Q^2 =
- 2 q^+ \, q^-$, such that, since $x>0$, then, due to \eq{xBj}, $q^-
>0$ and we have $q^+ <0$.

The $\xi$-integral in \eq{2-term3} is analyzed in Appendix~\ref{A},
where the $z$ and $z'$ integrals are also carried out.  The result is
(see \eq{R25})
\begin{align}\label{2-term4}
  \Pi_{ij}^{(2)} \, = \, \delta_{ij} \, \Lambda^2 \, S_\perp \,
  \frac{1152 \, N_c^2}{\pi^2} \, i \, \left( \frac{\mu}{a} \right)^2
  \frac{(q^-)^4}{|q^+| \, a \, Q^{10}} \, \int\limits_0^{\infty} \,
  \frac{d \, y}{(1+y)^{12}} \, y \, (1-y)^2 \left[ 1 + i \, q^+ \, a
    \, (1 + y) - e^{i \, q^+ \, a \, (1 + y)} \right].
\end{align}
If one performs $y$-integration in \eq{2-term4} one obtains an answer
expressed in terms of special functions. However this does not appear
to make the expression (\ref{2-term4}) more transparent: we will leave
it in the integral form. \eq{2-term4} is our exact result for the
transverse components of the hadronic tensor $\Pi_{ij}$ at the order
$\mu^2$.

To explicitly find corrections to the eikonal expression we expand
\eq{2-term4} in powers of $a$ and integrate over $y$ to obtain
\begin{align}\label{2-term5}
  \Pi_{ij}^{(2)} \, = \, \delta_{ij} \, \Lambda^2 \, S_\perp \,
  \frac{32 \, N_c^2}{7 \, \pi^2} \, i \, \left( \frac{\mu}{a}
  \right)^2 \, \frac{(q^-)^4}{Q^{10}} \, |q^+| \, a \, \left[ 1 + i \,
    \frac{2}{5} \, q^+ \, a - \frac{1}{8} \, (q^+ \, a)^2 + \ldots
  \right].
\end{align}
Using Eqs. (\ref{qa}), (\ref{q/a}), and (\ref{mq}), we rewrite
\eq{2-term5} in terms of $x$ and $Q^2$
\begin{align}\label{2-term6}
  \Pi_{ij}^{(2)} \, = \, \delta_{ij} \, \Lambda^2 \, S_\perp \,
  \frac{2 \, N_c^2}{7 \, \pi^2} \, i \, \frac{\Lambda^4 \,
    A^{1/3}}{Q^2 \, x^3} \, \left[ 1 - i \, \frac{2}{5} \, x \,
    A^{1/3} - \frac{1}{8} \, \left( x \, A^{1/3} \right)^2 + \ldots
  \right].
\end{align}
Finally, employing Eqs.~(\ref{Q1}) and (\ref{Q2}) we rewrite our
result (\ref{2-term6}) as
\begin{align}\label{2-term7}
  \Pi_{ij}^{(2)} \, = \, \delta_{ij} \, \Lambda^2 \, S_\perp \,
  \frac{8 \, N_c^2}{7 \, \pi^2} \, i \, \frac{\Lambda^2}{x^2} \,
  \frac{Q_1^2 (x,A)}{Q^2} \, \left[ 1 - i \, \frac{1}{10} \, \left(
      \frac{Q_2 (A)}{Q_1 (x,A)} \right)^2 - \frac{1}{128} \, \left(
      \frac{Q_2 (A)}{Q_1 (x,A)} \right)^4 + \ldots \right].
\end{align}

We now move on to the longitudinal components of the hadronic tensor.
\eq{n-term++} gives
\begin{align}\label{2-term1++}
  \Pi_{--}^{(2)} & \, = \, \frac{\Lambda^2 \, S_\perp}{a} \,
  \frac{N_c^2}{64 \, \pi^2} \, \left( \frac{\mu}{a} \right)^2 \, Q^4
  \, (q^-)^2 \, \int\limits_0^a d x^- \, \int\limits_0^a \, d y^- \,
  e^{i \, q^+ \,(x^- - y^-)} \, \int\limits_0^\infty \, d z \, z^5 \,
  K_0 (z \, Q) \notag \\ & \times \, \int\limits_0^\infty d z' \, I_0
  \left( z_< \, \sqrt{-i \, 2 \, q^- \, \partial_-} \right) \, K_0
  \left( z_> \, \sqrt{-i \, 2 \, q^- \, \partial_-} \right) \, z'^5 \,
  \, K_0 (z' \, Q) \, \delta (x^- - y^-).
\end{align}

The rest of evaluation proceeds along the same lines as for the
transverse components of $\Pi^{\mu\nu}$. Similar to \eq{2-term3} we
write
\begin{align}\label{2-term2++}
  \Pi_{--}^{(2)} & \, = \, \frac{\Lambda^2 \, S_\perp}{a} \,
  \frac{N_c^2}{64 \, \pi^2} \, \left( \frac{\mu}{a} \right)^2 \,
  \frac{Q^4 \, (q^-)^2}{|q^+|} \, \int\limits_{-\infty}^\infty \frac{d
    \, \xi}{2 \, \pi} \, \frac{1}{(1-\xi + i \, \epsilon)^2} \, \left[
    2 - e^{- i \, q^+ \, a \, (1 - \xi + i \, \epsilon)} - e^{i \, q^+
      \, a \, (1 - \xi + i \, \epsilon)} \right] \notag \\ & \times \,
  \int\limits_0^\infty \, d z \, z^5 \, K_0 (z \, Q) \,
  \int\limits_0^\infty d z' \, I_0 \left( z_< \, Q \, \sqrt{\xi - i \,
      \epsilon} \right) \, K_0 \left( z_> \, Q \, \sqrt{\xi - i \,
      \epsilon} \right) \, z'^5 \, \, K_0 (z' \, Q).
\end{align}
For evaluation of $z$-, $z'$- and $\xi$-integrals in \eq{2-term2++}
the reader is referred to Appendix \ref{A}. Using \eq{RL23} there we
write
\begin{align}\label{2-term3++}
  \Pi_{--}^{(2)} & \, = \, \Lambda^2 \, S_\perp \, \frac{32 \,
    N_c^2}{\pi^2} \, i \left( \frac{\mu}{a} \right)^2
  \frac{(q^-)^2}{|q^+| \, a \, Q^8} \int\limits_0^{\infty} \frac{d \,
    y}{(1+y)^{12}} \, (1 - 4 \, y + y^2)^2 \, \left[ 1 + i \, q^+ \, a
    \, (1 + y) - e^{i \, q^+ \, a \, (1 + y)} \right].
\end{align}
This is our final exact result for $\Pi_{--}$ at the order-$\mu^2$.

Expanding \eq{2-term3++} in the powers of $a$ yields
\begin{align}\label{2-term4++}
  \Pi_{--}^{(2)} & \, = \, \Lambda^2 \, S_\perp \, \frac{32 \,
    N_c^2}{35 \, \pi^2} \, i \left( \frac{\mu}{a} \right)^2
  \frac{(q^-)^2}{Q^8} \, |q^+| \, a \, \left[ 1 + i \, \frac{3}{8} \,
    q^+ \, a - \frac{1}{9} \, (q^+ \, a)^2 + \ldots \right],
\end{align}
or, in terms of Bjorken $x$ and $Q^2$,
\begin{align}\label{2-term5++}
  \Pi_{--}^{(2)} & \, = \, \Lambda^2 \, S_\perp \, \frac{2 \,
    N_c^2}{35 \, \pi^2} \, i \frac{Q^2}{(q^-)^2} \, \frac{\Lambda^4 \,
    A^{1/3}}{Q^2 \, x^3} \, \left[ 1 - i \, \frac{3}{8} \, x \,
    A^{1/3} - \frac{1}{9} \, \left( x \, A^{1/3} \right)^2 + \ldots
  \right],
\end{align}
and in terms of $Q_1$ and $Q_2$,
\begin{align}\label{2-term6++}
  \Pi_{--}^{(2)} & \, = \, \Lambda^2 \, S_\perp \, \frac{8 \,
    N_c^2}{35 \, \pi^2} \, i \frac{Q^2}{(q^-)^2} \,
  \frac{\Lambda^2}{x^2} \, \frac{Q_1^2 (x, A)}{Q^2} \, \left[ 1 - i \,
    \frac{3}{32} \, \left( \frac{Q_2 (A)}{Q_1 (x,A)} \right)^2 -
    \frac{1}{144} \, \left( \frac{Q_2 (A)}{Q_1 (x,A)} \right)^4 +
    \ldots \right].
\end{align}

One can see that the form of the hadronic tensor suggested in
\eq{Power3} is explicitly confirmed by our results in
Eqs.~(\ref{2-term7}) and (\ref{2-term6++})! The prefactors of the
square brackets of Eqs.~(\ref{2-term7}) and (\ref{2-term6++}) can also
be obtained from the eikonal expression (\ref{Pij4_eik}). For $Q^2 =
Q_2^2 (A)$ the second terms in the square brackets of
Eqs.~(\ref{2-term7}) and (\ref{2-term6++}) become parametrically
comparable to (and numerically much larger than) the leading-order
results given in \eq{1-term2}. As we noted above, this indicates the
breakdown of the eikonal formula (\ref{Pij4_eik}) at $Q^2 \sim Q_2^2
(A)$.


\subsection{Brief Summary of Our Results and Expressions for Structure Functions}
\label{sec:sum}

Let us briefly summarize the results of this Section. We have written
down exact general expressions (\ref{Pij3}) and (\ref{P++2}) for the
two independent components of the hadronic tensor $\Pi^{\mu\nu}$.
These expressions do not appear to be easy to evaluate in general
since they involve multiple iterations of the Green function operators
${\hat G}_z$ and ${\hat G}_z^L$. Instead we have employed two
approximations aimed at understanding the structure of the full
solution.

We first re-derived the components of the hadronic tensor in the
eikonal approximation
\cite{Hatta:2007he,Levin:2008vj,Mueller:2008bt,Avsar:2009xf} obtaining
\begin{subequations}\label{Pmn_tot}
\begin{align}
  \Pi^{ij}_{eik} (q^+, q^-, {\un q}=0) \, & = \, i \, \Lambda^2 \,
  S_\perp \, \frac{N_c^2}{16 \, \pi^2} \, \delta^{ij} \, \frac{Q^4}{x
    \, A^{1/3}} \int\limits_0^\infty \, d z \, z \, \left[ K_1 (Q \, z
    ) \right]^2 \, \left[ 1 - \exp \left( i \, Q^2 \, Q_1^2 (x, A) \,
      z^4
    \right) \right], \\
  \Pi^{++}_{eik} (q^+, q^-, {\un q}=0) \, & = \, i \, \Lambda^2 \,
  S_\perp \, \frac{N_c^2}{16 \, \pi^2} \, \frac{Q^2}{4 \, (q^-)^2} \,
  \frac{Q^4}{x \, A^{1/3}} \int\limits_0^\infty \, d z \, z \, \left[
    K_0 (Q \, z ) \right]^2 \, \left[ 1 - \exp \left( i \, Q^2 \,
      Q_1^2 (x, A) \, z^4 \right) \right].
\end{align}
\end{subequations}

We can re-write this result in terms of dimensionless structure
functions $F_1$ and $F_2$ defined by
\begin{align}\label{Fdef}
  F_1 (x, Q^2) \, = \, \frac{1}{2 \, \pi \, \Lambda^2} \ \mbox{Im} \,
  \Pi_1 (x, Q^2), \ \ \ \ \ F_2 (x, Q^2) \, = \, \frac{- p \cdot q}{2
    \, \pi \, \Lambda^2} \ \mbox{Im} \, \Pi_2 (x, Q^2)
\end{align}
where we replaced the conventional proton mass by the typical
transverse momentum in the shock wave $\Lambda$. For ${\un p} = {\un
  q} =0$ case considered here we have
\begin{align}\label{PPP}
  \Pi^{ij}_{eik} (q^+, q^-, {\un q}=0) \, = \, \delta^{ij} \, \Pi_1, \ 
  \ \ \ \ \Pi^{++}_{eik} (q^+, q^-, {\un q}=0) \, = \, \frac{Q^2}{4 \,
    (q^-)^2} \, \left[ - \Pi_1 + \frac{Q^2}{4 \, x^2} \, \Pi_2
  \right].
\end{align}
Combining Eqs.~(\ref{Pmn_tot}), (\ref{Fdef}) and (\ref{PPP}) we write
\cite{Hatta:2007he,Levin:2008vj,Mueller:2008bt,Avsar:2009xf}\footnote{Note
  that $N_c$-counting here agrees with the perturbative calculations
  of the DIS structure functions for partons in color-adjoint
  representation with the target nucleus made of nucleons with $N_c^2$
  valence partons each.}
\begin{subequations}\label{F2Leik}
\begin{align}
  F_2 (x, Q^2) \, & = \, S_\perp \, \frac{N_c^2}{16 \, \pi^3} \,
  \frac{Q^4}{A^{1/3}} \int\limits_0^\infty \, d z \, z \, \left[ K_1
    (Q \, z )^2 + K_0 (Q \, z )^2 \right] \, \text{Re} \, \left[ 1 -
    \exp \left( i \, Q^2 \, Q_1^2 (x, A) \, z^4 \right) \right], \label{F2eik} \\
  F_L (x, Q^2) \, & = \, S_\perp \, \frac{N_c^2}{16 \, \pi^3} \,
  \frac{Q^4}{A^{1/3}} \int\limits_0^\infty \, d z \, z \, \left[ K_0
    (Q \, z) \right]^2 \, \text{Re} \, \left[ 1 - \exp \left( i \, Q^2
      \, Q_1^2 (x, A) \, z^4 \right) \right],
\end{align}
\end{subequations}
where, as usual, $F_L = F_2 - 2 \, x \, F_1$. 

We have argued that the eikonal expressions (\ref{F2Leik}) apply only
for $Q^2 \gtrsim Q_1^2 (x,A)$. To demonstrate this explicitly we have
evaluated the hadronic tensor at the orders $\mu$ and $\mu^2$ going
beyond the eikonal approximation. Our above calculations can be
summarized as follows:
\begin{subequations}\label{Pi12}
  \begin{align}
    \Pi_1 (x, Q^2) \, = \, \Lambda^2 \, S_\perp \, \frac{N_c^2}{40 \,
      \pi^2} \, \frac{\Lambda^2}{x^2} \, \left[ 1 + i \, \frac{320}{7}
      \, \frac{Q_1^2 (x,A)}{Q^2} \, + \, \frac{32}{7} \, \frac{Q_2^2
        (A)}{Q^2} - i \, \frac{5}{14} \, \frac{Q_2^4 (A)}{Q^2 \, Q_1^2
        (x,A)} + \ldots
    \right], \\
    \Pi_2 (x, Q^2) \, = \, \Lambda^2 \, S_\perp \, \frac{4 \,
      N_c^2}{\pi^2} \, \frac{\Lambda^2}{Q^2} \, \left[ \frac{1}{24} +
      i \, \frac{72}{35} \, \frac{Q_1^2 (x, A)}{Q^2} + \frac{1}{5} \,
      \frac{Q_2^2 (A)}{Q^2} - i \, \frac{11}{720} \, \frac{Q_2^4
        (A)}{Q^2 \, Q_1^2 (x, A)} + \ldots \right].
  \end{align}
\end{subequations}
Clearly for $Q^2 \approx Q_2^2 (A)$ the third term in the brackets in
each of the equations (\ref{Pi12}) becomes comparable to the first
(leading-order) term in the series, thus generating an order-one
correction to Eqs.~(\ref{Pmn_tot}). Corrections to the structure
functions $F_2$ and $F_L$ result from the imaginary parts of $\Pi_1$
and $\Pi_2$ in \eq{Pi12}: at the order of the calculation shown in
\eq{Pi12} such corrections appear to stem only from the last terms in
the square brackets, which are always smaller than the 2nd terms
contributing to the structure functions. However, it is clear that an
order-$\mu^3$ calculation would generate {\sl imaginary} terms
$\propto i \, Q_1^2 \, Q_2^2/Q^4$ in the square brackets of $\Pi_1$
and $\Pi_2$, which would become comparable to leading large-$Q^2$
contributions to the structure functions for $Q^2 \approx Q_2^2 (A)$,
generating corrections to (\ref{F2Leik}). Such order-$\mu^3$
calculation, while conceptually straightforward, is technically rather
involved. We have verified that the terms $\propto i \, Q_1^2 \,
Q_2^2/Q^4$ do indeed arise in such calculation: the determination of
the exact numerical prefactors in front of such terms does not seem to
be important for the conceptual conclusion about the breakdown of the
eikonal formulas (\ref{F2Leik}) at $Q^2 \approx Q_2^2 (A)$ which we
draw here.

However, the applicability region of Eqs.~(\ref{F2Leik}) is not simply
$Q^2 \gg Q_2^2 (A)$.  In fact, as was detailed in Sec. \ref{appl},
non-eikonal corrections to the higher-order terms in the eikonal
series lead to the applicability region of the eikonal approach being
reduced to $Q^2 \gtrsim Q_1^2 (x,A)$, as such corrections become
important at scales arbitrary close to (but smaller than) $Q_1^2
(x,A)$. At the moment we can not asses the net size and the effect of
such corrections: this may require knowing the exact solution of the
problem (i.e., the exact solution of Eqs. (\ref{Maxwell_all})).


\section{Discussion of Momentum Scales in DIS}
\label{MS}

Above we have shown that $R$-current DIS on a shock wave of finite
longitudinal extent at strong 't Hooft coupling is described by two
momentum stales, $Q_1$ and $Q_2$. This seems to be natural since the
finite-size shock wave is described by two dimensionful scales: $\mu$
and $a$. 

Our conclusion also appears to be in qualitative agreement with the
calculation performed in \cite{Albacete:2008ze}: there DIS process on
a shock wave was modeled by a quark--anti-quark dipole scattering on a
shock wave. Indeed in QCD this is how DIS process takes place: virtual
photon splits into a quark--anti-quark pair which then scatters on a
target proton or nucleus (see e.g.
\cite{Kovchegov:1999yj,Kovchegov:1999ua} and references therein). In
\cite{Albacete:2008ze} the dipole--shock wave scattering was described
by calculating an expectation value of a fundamental Wilson loop in
the shock wave background. For a shock wave of the type (\ref{nuc})
which does not have any transverse coordinate dependence, the
resulting forward dipole-target scattering amplitude $N(r,s)$ was a
function of the transverse dipole size $r$ and the center-of-mass
energy $s$. In order to obtain the DIS structure functions one has to
convolute $N(r,s)$ with the light-cone wave function of a virtual
photon splitting into $q\bar q$ pair, which results in $r$ being dual
to $1/Q$ (see e.g. \cite{Nikolaev:1990ja,Kovchegov:2009yj}). The
dipole amplitude $N(r,s)$ from \cite{Albacete:2008ze} had two momentum
scales associated with it. One scale was defined by the condition
$(\mu/a) \, r^4 =$~const and indicated the $q\bar q$ separation $r$ at
which the classical string solution became complex-valued. To
translate this scale into $Q^2$ and $x$ variables we note that the
calculation in \cite{Albacete:2008ze} was done in the rest frame of
the dipole.  Taking into account Lorentz properties of $\mu/a$ we see
that to generalize this condition we should replace $r^4$ by
$(q^-)^2/Q^6$ (as the appropriately transforming projectile-related
parameters), such that this scale, which we label $Q_3$, is defined by
the condition
\begin{align}
  \frac{\mu}{a} \, \frac{(q^-)^2}{(Q_3)^6} \, = \, \text{const}
\end{align}
leading to 
\begin{align}\label{Q3}
  Q_3 (x) \, \sim \, \frac{\Lambda}{x} \, \sim \, \frac{Q_1 (x,
    A)^2}{Q_2 (A)}.
\end{align}
The second scale describing DIS obtained in \cite{Albacete:2008ze} was
the scale $Q_2 (A)$. In complete analogy with perturbative QCD
calculations, defining saturation scale by $N (1/Q_s, s) = 1/2$ (half
of the black disk limit of $N=1$) \cite{Albacete:2007yr}, the
calculation in \cite{Albacete:2008ze} obtained $Q_s \approx Q_2 (A)$
(see also \cite{Dominguez:2008vd}). It was also observed that the
dipole amplitude $N(r,s)$ became independent of energy/Bjorken $x$ at
high energy for a broad range of values of $r$ both inside and outside
of the saturation region.

We see that the two scales $Q_1$ and $Q_2$ we have obtained above were
also present in the calculation of \cite{Albacete:2008ze}. In the
small-$x$ regime, when $x \, A^{1/3} \ll 1$, the lower scale found in
\cite{Albacete:2008ze} was $Q_2$ which corresponded to the saturation
scale. The larger scale $Q_3$ from \eq{Q3} can also be understood: in
\cite{Albacete:2008ze} the string solution considered was static,
which is strictly speaking only valid for a shock wave of infinite
extent.  Hadronic tensor for such shock wave can be obtained from our
exact equations (\ref{Pij3}) and (\ref{P++2}) by taking $a \rightarrow
\infty$ limit while keeping $\mu/a$ fixed: this would simply remove
theta-functions, making the series in Eqs.~(\ref{Pij3}) and
(\ref{P++2}) a power series in $(\mu/a) \, (q^-)^2/Q^6$. The problem
now is described by only one scale --- the scale $Q_3$ from \eq{Q3}.

To reconcile this single-scale result with our two-scale conclusion
above one could argue following \cite{Hatta:2007he} that for DIS on an
infinite-extent shock wave what matters is the size of the interaction
region between the $R$-currents and the shock wave. One should
therefore replace the shock wave longitudinal width $a \sim
A^{1/3}/p^+$ by the typical longitudinal separation between points
$x^-$ and $y^-$ in the $R$-current correlator (\ref{Pdef}): the latter
is $\sim 1/|q^+| = 1/(x \, p^+)$.  Hence one has to replace
\begin{align}\label{Ax}
  A^{1/3} \rightarrow \frac{1}{x}.
\end{align}
Under such replacement both scales $Q_1$ and $Q_2$ become equal to
$Q_3$ and the problem becomes single-scale. This is also why the
saturation scale $Q_s = Q_2$ found in \cite{Albacete:2008ze} is in
agreement with the results of \cite{Hatta:2007he,Hatta:2007cs} for the
infinite medium: under the substitution (\ref{Ax}) one has $(Q_2)^2
= \Lambda^2 A^{2/3} \rightarrow \Lambda^2 /x^2 = (Q_3)^2$. (The
discrepancy by a factor of $\sqrt{\lambda}$ with $\lambda$ the 't
Hooft coupling that we mentioned above still remains indeed: we can
not explain it at the moment.)

Let us now return to the finite-extent shock waves.  While both our
above analysis and the calculations of \cite{Albacete:2008ze} have the
same conclusion about DIS at strong coupling being a two-scale
problem, one may still worry about the physical interpretation of the
scales $Q_1$ and $Q_2$ we found. The calculation in
\cite{Albacete:2008ze} considered DIS on a finite-size shock wave but,
as a first approximation, employed the static limit of the string
configuration, strictly speaking valid for an infinite shock wave
only: the question whether in scattering on a large but finite shock
wave the string has enough time to quickly settle onto its static
configuration remains to be answered (see \cite{Mueller:2008bt} for
the analysis of the problem for a thin shock wave). Moreover, building
on the analogy with the complex trajectory method in Quantum Mechanics
the classical string solutions found in \cite{Albacete:2008ze} were
analytically continued into the complex-valued domain. Justification
of such a procedure may be needed in the string theory context.
Therefore we will try to discuss the possible physical meanings of
$Q_1$ and $Q_2$ with an open mind while temporarily ignoring the
pre-existing results of \cite{Albacete:2008ze}.

Indeed the physical meaning of the scales $Q_1$ and $Q_2$ would have
probably been more manifest if the exact result for $F_2$ structure
function was known. Instead we have the eikonal expression
(\ref{F2eik}) due to
\cite{Hatta:2007he,Levin:2008vj,Mueller:2008bt,Avsar:2009xf}, which is
valid for $Q^2 \gtrsim Q_1^2 (x,A)$. At large $Q^2 \gg Q_1^2 (x,A)$ it
gives \cite{Mueller:2008bt,Avsar:2009xf,Levin:2008vj}
\begin{align}\label{Qlarge}
  F_2 (x, Q^2 ) \big|_{Q^2 \gg Q_1^2 (x,A)} \, \approx \, S_\perp \,
  \frac{18 \, N_c^2}{35 \, \pi^3} \, \frac{\Lambda^4 \, A^{1/3}}{x^2 \, Q^2},
\end{align}
while at small $Q^2$ it reduces to \cite{Avsar:2009xf,Levin:2008vj}
\begin{align}\label{Qsmall}
  F_2 (x, Q^2 ) \big|_{Q^2 \ll Q_1^2 (x,A)} \, \approx \, S_\perp \,
  \frac{N_c^2}{64 \, \pi^3} \, \frac{Q^2}{A^{1/3}} \, \ln \frac{Q_1^2
    (x,A)}{Q^2},
\end{align}
though it is not clear how reliable \eq{Qsmall} is in light of $Q^2
\gtrsim Q_1^2 (x,A)$ applicability constraint of \eq{F2eik}. We will
proceed under assumption that Eqs.~(\ref{Qlarge}) and (\ref{Qsmall})
are qualitatively correct, i.e., that $F_2 (x, Q^2 )$ has a maximum at
$Q^2 \approx Q_1^2 (x,A)$ and it decreases as $Q^2$ becomes either
larger or smaller than $Q_1^2 (x,A)$.

The exact effect of the scale $Q_2 (A)$ on $F_2$ is not clear from the
above calculations and it appears that to clarify it one needs to
solve the problem exactly. In the meantime we argue that
Eqs.~(\ref{Pi12}) indicate that the structure functions would change
quite significantly at $Q = Q_2 (A)$. We can only guess the exact
effect of $Q_2$ on $F_2$. If we believe that $F_2$ already decreases
with decreasing $Q^2$ for $Q < Q_1$, as seems to follow from
\eq{Qsmall} which we choose to believe at least at the qualitative
level, and combine this with the fact that, on general grounds, $F_2$
should go to $0$ for $Q^2 \rightarrow 0$, we conclude that it is
probable that for $Q < Q_2$ the structure function $F_2$ would
continue to decrease with decreasing $Q^2$, probably decreasing faster
than it was for $Q_2 < Q < Q_1$. The sketch of our guess/tentative
understanding of $F_2 (x, Q^2)$ is shown in \fig{f2}.

\FIGURE{\includegraphics[width=11cm]{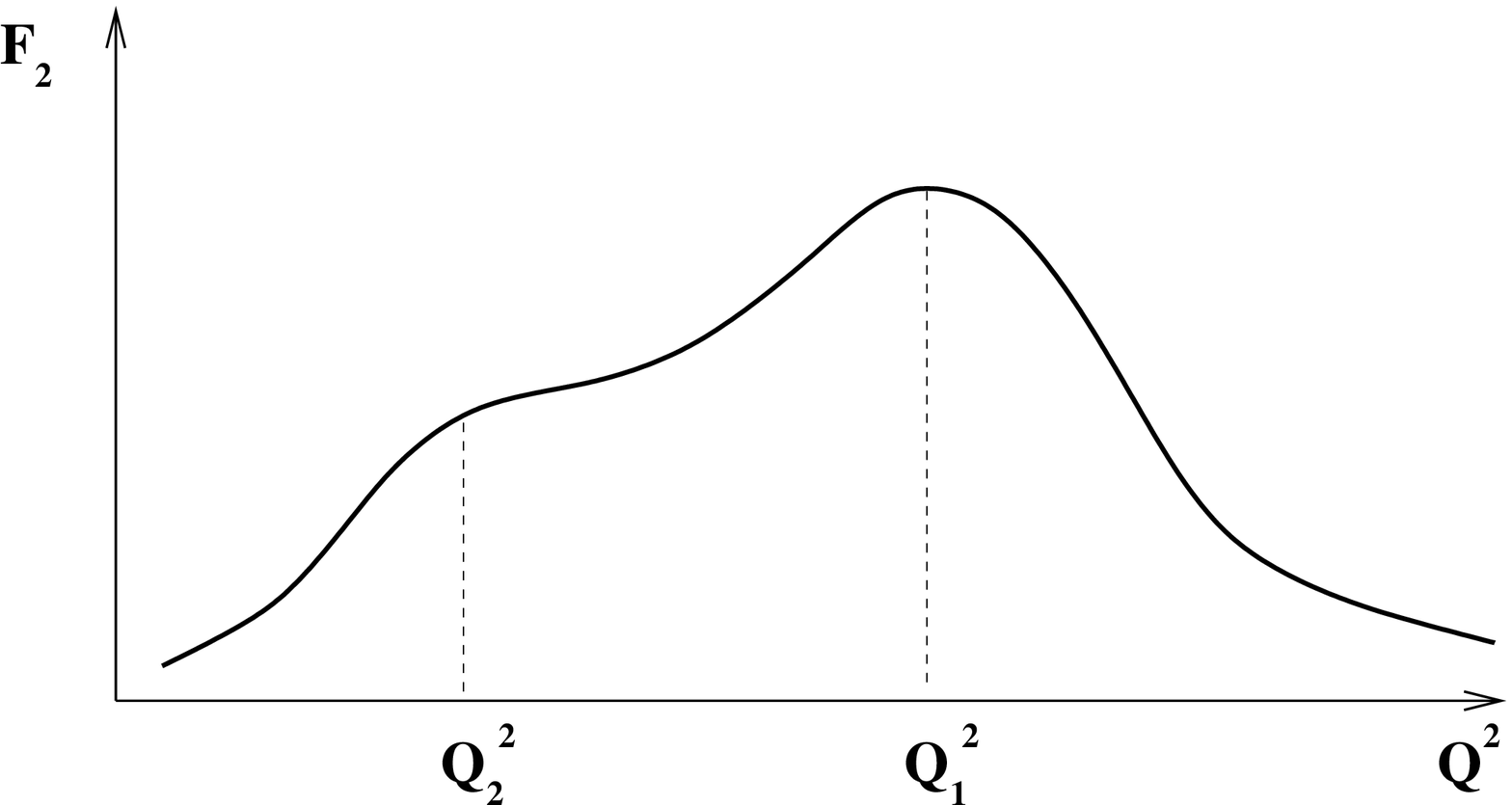}
  \caption{A sketch of the $F_2$ structure function of $R$-current DIS 
    at strong coupling as a function of $Q^2$ based on our
    understanding/guess of the exact AdS/CFT prediction (see text).}
  \label{f2}
}

Assuming that the sketch in \fig{f2} accurately represents the
structure function $F_2$ given by the exact solution of the
$R$-current scattering problem in AdS/CFT, we propose the following
three possible interpretations of the physical meaning of the scales
$Q_1$ and $Q_2$.

~\\

1. The first, and, in our view, the most probable option, stems from
comparing $F_2$ structure function in \fig{f2} to what one has in QCD
at small coupling and/or to the actual data reported by DIS
experiments. The prediction of CGC/saturation physics is that the
$F_2$ structure function scales as
\cite{Kovchegov:1999ua,Mueller:2002zm,Levin:1999mw,Iancu:2002tr} (for
a review see \cite{Jalilian-Marian:2005jf})\footnote{The power of
  $Q^2$ in \eq{Qg} is given by the fixed-coupling approximation. This
  power changes when running coupling corrections are included
  \cite{Albacete:2007yr,Albacete:2004gw}. For our purposes we only
  need the power to be positive and smaller than 1, which is true at
  both the fixed and running coupling.}
\begin{subequations}\label{CGC}
\begin{align}
  F_2^{CGC} \, \propto \, Q^2 , \ \ \ \ \ Q^2 \ll Q_s^2 \label{Q22} \\
  F_2^{CGC} \, \propto \, (Q^2)^{0.628} , \ \ \ \ \ Q^2 \gtrsim Q_s^2.
  \label{Qg}
\end{align}
\end{subequations}
It is important to note that in the small-$x$ CGC/saturation physics
framework $F_2$ structure function never decreases with increasing
$Q^2$! Therefore it may seem hard to reconcile the decrease of $F_2$
with $Q^2$ shown in \eq{Qlarge} with small-$x$ CGC/saturation physics.
Moreover, if one remembers that in weakly-coupled QCD $F_2$ is given
by the sum of quark distributions $x q (x, Q^2)$ over all flavors the
decrease with $Q^2$ may seem even more puzzling: if we use the
standard (albeit somewhat simplified) interpretation of $x q (x, Q^2)$
as the number of quarks at a given value of Bjorken $x$ with
transverse momenta $k_T \le Q$, then it would appear that $x q (x,
Q^2)$ along with $F_2 (x, Q^2)$ can never decrease with $Q^2$, since
the number of quarks with $k_T \le Q$ can only increase with $Q^2$.

However such arguments are not entirely correct. It relies on a simple
perturbative relation between $F_2 (x, Q^2)$ and distribution
functions, which may be modified at strong coupling. On top of
that\footnote{We would like to thank Genya Levin for pointing out this
  argument to us.}, at large-$Q^2$, when the
Dokshitzer--Gribov--Lipatov--Altarelli--Parisi (DGLAP) evolution
\cite{Dokshitzer:1977sg,Gribov:1972ri,Altarelli:1977zs} is important,
the relation between, say, the gluon distribution function $xG (x,
Q^2)$ and the unintegrated gluon distribution $\phi$ is not simply
\begin{align}\label{xGphi}
  xG (x, Q^2) \, = \, \int\limits^{Q^2} \, d k_T^2 \ \phi (x, k_T^2)
\end{align}
but, instead, is given by
\cite{Dokshitzer:1978hw,Catani:1989sg,Kimber:2001sc,Kimber:1999xc,Khoze:2000cy}
\footnote{Of course \eq{ff} should get substantially
  modified and ceases to be valid also at low-$Q^2$ inside the
  saturation region ($Q < Q_s$) due to multiple-rescatterings,
  non-linear evolution, and other higher-twist corrections.}
\begin{align}\label{ff}
  xG (x, Q^2) \, = \, \int\limits^{Q^2} \, d k_T^2 \ \phi (x, k_T^2,
  Q^2).
\end{align}
As usual for distribution functions, $Q^2$ is the renormalization
scale: in the spirit of the leading logarithmic approximation we put
it as the upper cutoff on the $k_T^2$ integral in \eq{ff}.  \eq{ff}
shows that when one goes beyond the leading logarithmic small-$x$
evolution approximation, and includes DGLAP evolution
\cite{Dokshitzer:1977sg,Gribov:1972ri,Altarelli:1977zs} as well, the
unintegrated gluon distribution $\phi (x, k_T^2, Q^2)$ itself becomes
a function of $Q^2$. It implies that the highest transverse momentum
of a ``real'' parton in the proton's wave function is $k_T$, while the
wave function is evolved using DGLAP to the scale $Q^2$, such that the
evolution from $k_T^2$ to $Q^2$ is due to virtual corrections to DGLAP
only, resulting in a form-factor in the definition of $\phi (x, k_T^2,
Q^2)$ (see
\cite{Catani:1989sg,Kimber:2001sc,Kimber:1999xc,Khoze:2000cy} for more
details and for the definition of $\phi (x, k_T^2, Q^2)$). Now, as
$Q^2$ increases, the unintegrated distribution function $\phi (x,
k_T^2, Q^2)$ decreases, as the probability of no real gluon emissions
between $k_T^2$ and $Q^2$ decreases with increasing $Q^2$. It is thus
possible that at large enough $Q^2$ this decrease with $Q^2$ in $\phi$
would dominate in \eq{ff}, resulting in the gluon distribution
function $xG (x, Q^2)$ decreasing with $Q^2$.  (The same argument can
be applied to quark distributions.)

It is important to point out that if one wants to interpret $\phi (x,
k_T^2, Q^2)$ in \eq{ff} as the number of gluons with transverse
momentum $k_T$, this number would depend on the momentum of the probe
(or, equivalently, on the renormalization scale) $Q$. The reason
behind this $Q$-dependence is that $\phi (x, k_T^2, Q^2)$ really gives
the number of gluons at $k_T$ with the condition that there are no
gluons with higher transverse momenta in the hadronic wave function.
It appears to be impossible in general to define unintegrated gluon
distribution independent of $Q^2$, which would simply give the number
of gluons at $k_T$ without any exclusive conditions. Thus the
probabilistic interpretation of the gluon distribution $xG (x, Q^2)$
as the number of gluons with $k_T \le Q$ is not valid once the full
DGLAP evolution is included. (Again the same applies to quark
distributions.) This is why the falloff of $F_2$ with $Q^2$ presents
no contradiction.

To visualize how a distribution function (and therefore a structure
function) may decrease with $Q^2$ and to determine at what $Q^2$ these
functions start decreasing let us consider a simple but realistic toy
model.\footnote{We are grateful to Genya Levin for this argument as
  well.}  Take the gluon distribution given by the solution of the
leading-logarithmic fixed-coupling DGLAP evolution equation:
\begin{align}\label{xGtoy}
  x G (x, Q^2) \, = \, \int\limits_{b-i \, \infty}^{b+ i \, \infty} \,
  \frac{d \omega}{2 \, \pi \, i} \ x^\omega \, \left(
    \frac{Q^2}{Q_0^2} \right)^{\gamma_{GG} (\omega)} \, G_\omega (Q_0^2).
\end{align}
Here $b$ is an arbitrary real number and $Q_0$ is the initial scale of
DGLAP evolution. For simplicity we assume that there are no quarks in
the toy theory we consider. We also assume a particularly simple toy
form of the gluon-gluon splitting function \cite{Ellis:1993rb}
\begin{align}\label{GG}
  \gamma _{GG} (\omega) \, = \, \frac{\as \, N_c}{\pi} \, \left(
    \frac{1}{\omega} -1 \right).
\end{align}
This splitting function has the correct residue of the small-$x$ pole
at $\omega =0$. The term $(-1)$ in the parenthesis of \eq{GG} mimics
all the non--small-$x$ terms in the actual splitting function. It also
makes sure that the momentum sum rule
\begin{align}\label{sum}
  \gamma _{GG} (1) \, = \, 0
\end{align}
is satisfied.

At small-$x$ and large-$Q^2$ the integral in \eq{xGtoy} can be
evaluated in the saddle-point approximation with the saddle point at
\begin{align}
  \omega_{sp} \, = \, \sqrt{\frac{\as \, N_c}{\pi} \, \frac{\ln (Q^2 /
      Q_0^2)}{\ln (1/x)}}
\end{align}
and the gluon distribution given approximately by
\begin{align}
  x G (x, Q^2) \, \sim \, x^{\omega_{sp}} \, \left( \frac{Q^2}{Q_0^2}
  \right)^{\gamma_{GG} (\omega_{sp})}.
\end{align}
This distribution function is a decreasing function of $Q^2$ for
$\gamma_{GG} (\omega_{sp}) <0$, which means $\omega_{sp} > 1$.
Therefore the gluon distribution decreases with $Q^2$ for
\begin{align}\label{decr}
  Q^2 \, > \, Q_{decr}^2 \, \equiv \, Q_0^2 \, \left( \frac{1}{x}
  \right)^\frac{\pi}{\as \, N_c}.
\end{align}
This is indeed a very large scale for small-$x$, but for larger-$x$ it
becomes small enough for decrease of $F_2$ with $Q^2$ to be seen
experimentally at HERA. Note that at large 't Hooft coupling the scale
in \eq{decr} is not necessarily large.

\eq{decr} illustrates a known fact that at very large $Q^2$
distribution functions (and, therefore, structure functions) do
decrease with $Q^2$ even in the perturbative picture. Therefore,
combining this result with Eqs.~(\ref{CGC}) we now see that in
perturbative QCD the $F_2$ structure function looks qualitatively as
shown in \fig{f2} if we identify the scale $Q_2$ with the saturation
scale $Q_s$ and the scale $Q_1$ with the scale $Q_{decr}^2$ from
\eq{decr} at which $F_2$ starts falling off with $Q^2$.

Therefore our first guess at the physical meaning of $Q_2$ and $Q_1$
is to identify them with $Q_s$ and $Q_{decr}$ correspondingly. In
\cite{Avsar:2009xf,Hatta:2007he,Hatta:2007cs} it is shown that the
scale $Q_1$ is essential for satisfying the momentum sum rule: this
seems to confirm our conclusion since $Q_{decr}$ results from
satisfying the same momentum sum rule of \eq{sum}. It is possible that
at strong 't Hooft coupling, just like at small coupling $\as$, energy
conservation effects come in at a different $Q^2$-scale from
unitarization effects.

2. The second interpretation of $Q_2$ and $Q_1$ we propose is to leave
the interpretation of $Q_2$ as the saturation scale, but to suggest
that $Q_1$ is the extended geometric scaling scale $k_{geom}$
\cite{Iancu:2002tr,Levin:1999mw,Stasto:2000er}. Extended geometric
scale $k_{geom}$ is the scale such that for $Q < k_{geom}$ the
structure functions are functions of $Q/Q_s$ only
\cite{Iancu:2002tr,Levin:1999mw}. In CGC usually $k_{geom} > Q_s$
\cite{Iancu:2002tr}, which supports our hypothesis here. Also, if we
accept \eq{Qsmall} as being at least qualitatively correct for the
exact AdS/CFT prediction for $F_2$, we can see that $F_2$ is almost
completely $x$-independent below $Q_1$, and probably can be written as
a function of $Q/Q_2$, which also supports the suggestion that $Q_1$
could be $k_{geom}$, since in CGC $F_2$ is a function of $Q/Q_s$ for
$Q < k_{geom}$ \cite{Iancu:2002tr}. Indeed the relation between
$k_{geom}$ and $Q_s$ has to be modified at strong coupling in
comparison to the weak-coupling CGC result
\cite{Iancu:2002tr,Kharzeev:2003wz,Jalilian-Marian:2005jf}.

The main problem with this scenario is that in CGC for $Q > k_{geom}$
the structure function $F_2$ keeps increasing with $Q$
\cite{Iancu:2002tr,Jalilian-Marian:2005jf}, while in AdS/CFT
calculations one obtains $F_2$ decreasing with $Q$ for $Q > Q_1$ as
one can see from \eq{Qlarge} and from \fig{f2}. Therefore our second
hypothesis does not seem to be in agreement with the shape of the plot
in \fig{f2} for $Q > Q_1$, which makes it somewhat less compelling than
the first one.

3. Finally one may accept the viewpoint advocated in
\cite{Mueller:2008bt,Avsar:2009xf,Hatta:2007he,Hatta:2007cs} and
identify $Q_1$ with the saturation scale. Indeed the similarity
between Eqs.~(\ref{Q22}) and (\ref{Qsmall}) seems to suggest that this
is correct. However, as we argued above, \eq{Qsmall}, while obtained
by eikonal methods, lies outside the region of applicability of the
eikonal approximation and should be questioned. Also in CGC the
structure function $F_2$ continues growing with $Q^2$ for $Q > Q_s$,
as one can see from \eq{Qg}, in disagreement with \eq{Qlarge}, casting
more doubt on this third possible scenario. One should also mention
that $x$-independence of structure functions was observed at large
coupling in \cite{Albacete:2008ze} for $Q > Q_s$: therefore
$x$-independence of \eq{Qsmall} may not yet signal saturation.

An important question remains regarding the physical role of the scale
$Q_2$. In traditional CGC literature there are no important scales
below $Q_s$. One may speculate that $Q_2$ may be the scale at which
other higher twist effects, such as pomeron loops, may become
important (see e.g. \cite{Kovchegov:1999yj}). While possible in
principle we believe further research is needed to test this
assumption. Pomeron loops are suppressed by powers of $A$, while the
scale $Q_2$ does not have any $A$-suppression compared to $Q_1$,
exhibiting the opposite $A$-enhancement. In principle, until the exact
solution of the problem is found, it may also be possible that nothing
of physical importance happens at the scale $Q_2$, though such
conclusion is hardly likely, since a whole class of terms becomes
important at this scale, as one can see from \eq{Power3}. The scale
$Q_2$ is known to play an important role in heavy ion collisions
modeled in AdS/CFT: as was shown in
\cite{Albacete:2008vs,Albacete:2009ji} in a strongly-coupled collision
shock waves stop at the light-cone time $x^+_{stop} \sim 1/Q_2$ in the
center-of-mass frame.  It is probable that the scale that determines
the stopping time in a shock wave collision should play some role in
DIS as well. \\

Indeed an exact solution of the $R$-current DIS problem is needed to
conclude whether one (if any) of the above-listed possibilities is
correct.  Unfortunately an exact analytic evaluation of
Eqs.~(\ref{Pij3}) and (\ref{P++2}) appears to be a rather difficult
problem at present.


\acknowledgments

The author would like to thank Bo-Wen Xiao for explaining to him the
essential steps in computation of the $R$-current correlators using
AdS/CFT correspondence, Genya Levin for very useful and informative
discussions about the important scales in DIS, and Dionysis
Triantafyllopoulos for clarifying important details of
\cite{Avsar:2009xf}.

This research is sponsored in part by the U.S. Department of Energy
under Grant No. DE-FG02-05ER41377.



\appendix

\renewcommand{\theequation}{A\arabic{equation}}
  \setcounter{equation}{0}
\section{Some useful integrals}
\label{A}

We start by integrating over $\xi$ in \eq{2-term3}, namely we need to
find 
\begin{align}\label{R21}
  R_2 \, \equiv & \, \int\limits_{-\infty}^\infty \frac{d \, \xi}{2 \,
    \pi} \, \frac{1}{(1-\xi + i \, \epsilon)^2} \, \left[ 2 - e^{i \,
      q^+ \, a \, (1 - \xi + i \, \epsilon)} - e^{- i \, q^+ \, a \,
      (1 - \xi + i \, \epsilon)} \right] \notag \\ & \times \,
  \int\limits_0^\infty \, d z \, z^5 \, K_1 (z \, Q) \,
  \int\limits_0^\infty d z' \, I_1 \left( z_< \, Q \, \sqrt{\xi - i \,
      \epsilon} \right) \, K_1 \left( z_> \, Q \, \sqrt{\xi - i \,
      \epsilon} \right) \, z'^5 \, \, K_1 (z' \, Q).
\end{align}
Using the series representations of the modified Bessel functions
\begin{subequations}\label{IKser}
\begin{align}
  I_1 (z) \, & = \, \sum\limits_{m=0}^\infty \, \frac{1}{\Gamma (m) \,
    \Gamma (m+1)} \, \left( \frac{z}{2} \right)^{2 \, m -1}  \\
  K_1 (z) \, & = \, \sum\limits_{m=0}^\infty \, \frac{1}{\Gamma (m) \,
    \Gamma (m+1)} \, \left( \frac{z}{2} \right)^{2 \, m -1} \, \left[
    \ln \left( \frac{z}{2} \right) - \frac{1}{2 \, m} - \psi (m)
  \right]
\end{align}
\end{subequations}
we see that $I_1 \left( z_< \, Q \, \sqrt{\xi - i \, \epsilon} \right)
\, K_1 \left( z_> \, Q \, \sqrt{\xi - i \, \epsilon} \right)$ from
\eq{R21} has a branch cut discontinuity for $\xi \in ( -\infty + i \,
\epsilon, 0+ i \, \epsilon]$. The complex structure of the integrand
in \eq{R21} is depicted in \fig{compl}. The integrand has a branch cut
we have just mentioned, along with a possible pole at $\xi = 1 + i \,
\epsilon$. While strictly speaking there is no pole at $\xi = 1 + i \,
\epsilon$ in the full expression in \eq{R21}, individual terms in the
square brackets in \eq{R21} lead to contributions to the integrand
containing this pole.

\FIGURE{\includegraphics[width=11cm]{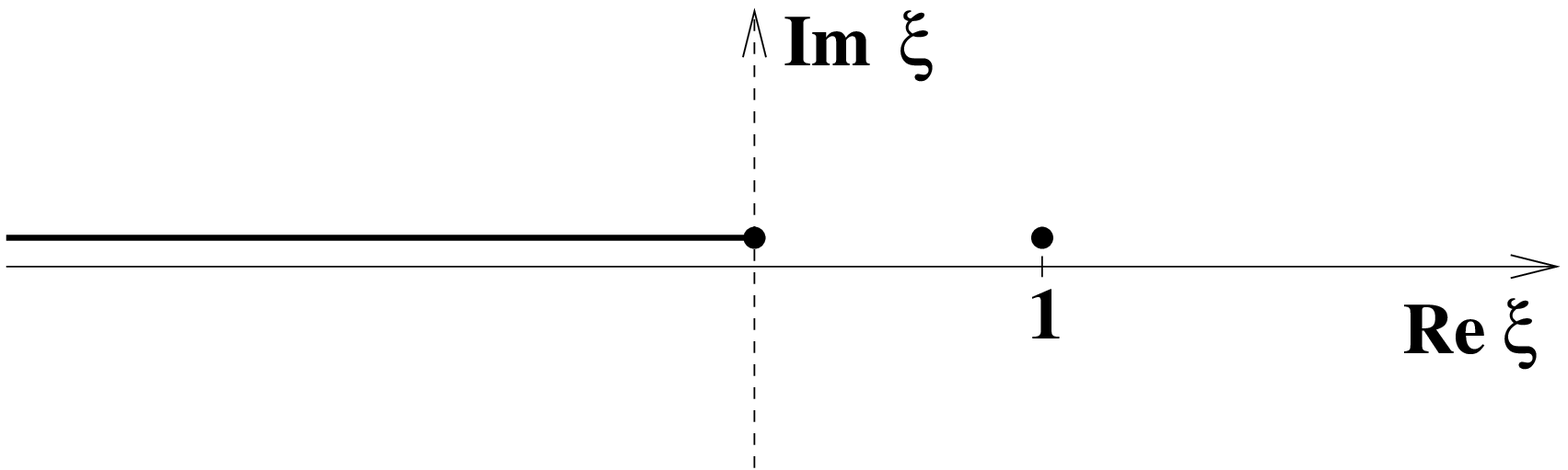}
  \caption{The complex $\xi$-plane structure of the integrands of 
   Eqs.~(\ref{R21}) and (\ref{RL21}) (see text).}
  \label{compl}
}

Since $q^+ <0$, the last term in the square brackets of \eq{R21}
demands that the $\xi$-integration contour be closed in the lower
half-plane: since there are no poles or branch cuts in the lower
half-plane, we can discard this term. For the first two terms in the
square brackets of \eq{R21} we have to close the integration contour
in the upper half-plane (for the very first term the direction of
contour closing is actually dictated by the large-argument
asymptotics of the modified Bessel functions). Picking up the pole at
$\xi = 1 + i \, \epsilon$ and wrapping the contour around the branch
cut yields
\begin{align}\label{R22}
  R_2 \, = & \, \frac{i}{2} \int\limits_{-\infty}^0 \, \frac{d \,
    \xi}{(1-\xi)^2} \, \left[ 2 - e^{i \, q^+ \, a \, (1 - \xi)}
  \right] \left[ \int\limits_0^\infty \, d z \, z^5 \, K_1 (z \, Q) \,
    J_1 \left( z \, Q \, \sqrt{-\xi} \right) \right]^2 + i \,
  \frac{d}{d \xi} \Bigg\{ \left[ 2 - e^{i \, q^+ \, a \,
      (1 - \xi)} \right] \notag \\
  & \times \, \int\limits_0^\infty \, d z \, z^5 \, K_1 (z \, Q) \,
  \int\limits_0^\infty d z' \, I_1 \left( z_< \, Q \, \sqrt{\xi}
  \right) \, K_1 \left( z_> \, Q \, \sqrt{\xi} \right) \, z'^5 \, \,
  K_1 (z' \, Q) \Bigg\} \Bigg|_{\xi = 1}
\end{align}
where we have used
\begin{align}
  I_1 \left( z_< \, Q \, \sqrt{\xi - i \, \epsilon} \right) \, K_1
  \left( z_> \, Q \, \sqrt{\xi - i \, \epsilon} \right) - I_1 \left(
    z_< \, Q \, \sqrt{\xi + i \, \epsilon} \right) \, K_1 \left( z_>
    \, Q \, \sqrt{\xi + i \, \epsilon} \right) \notag \\ \, = \, \pi
  \, i \, \theta (-\xi) \, J_1 \left( z \, Q \, \sqrt{-\xi} \right)
  \, J_1 \left( z' \, Q \, \sqrt{-\xi} \right)
\end{align}
which can be inferred from Eqs.~(\ref{IKser}).

The first term on the right-hand-side of \eq{R22} is straightforwardly
evaluated as
\begin{align}\label{R23}
  & \, \frac{i}{2} \int\limits_{-\infty}^0 \, \frac{d \,
    \xi}{(1-\xi)^2} \, \left[ 2 - e^{i \, q^+ \, a \, (1 - \xi)}
  \right] \left[ \int\limits_0^\infty \, d z \, z^5 \, K_1
    (Q \, z ) \, J_1 \left( z \, Q \, \sqrt{-\xi} \right) \right]^2 \notag \\
  & = \, \frac{i}{2} \, \frac{192^2}{Q^{12}} \,
  \int\limits_{-\infty}^0 \, \frac{d \, \xi}{(1-\xi)^2} \, \left[ 2 -
    e^{i \, q^+ \, a \, (1 -
      \xi)} \right] \left[ \frac{\sqrt{-\xi} \, (1+\xi)}{(1-\xi)^5} \right]^2  \notag \\
  & = \, \frac{i}{2} \, \frac{192^2}{Q^{12}} \, \int\limits_0^{\infty}
  \, \frac{d \, y}{(1+y)^{12}} \, y \, (1-y)^2 \, \left[ 2 - e^{i \,
      q^+ \, a \, (1 + y)} \right],
\end{align}
where $y = -\xi$.  The second term on the right-hand-side of \eq{R22}
is a linear polynomial in $a$,
\begin{align}\label{R24}
  \# + \#' \, a,
\end{align}
with the coefficients depending on $Q$. We know from the eikonal
approximation (see \eq{eik_long1}) that $R_2$, if expanded in a series
in the powers of $a$, should start at the order $a^2$. The same can be
inferred from \eq{R21}, though we note that a power-series in $a$
expansion in the integrand there gives finite results only at the
order $a^2$, not allowing to learn anything about higher powers of
$a$.

Requiring that the series in powers of $a$ for $R_2$ starts from $a^2$
along with \eq{R24} shows that the second term on the right-hand-side
of \eq{R22} simply cancels the constant and linear in $a$ terms in the
first term on the right-hand-side of \eq{R22}. (We have also checked
by explicit numerical integration that this is true.) Adding extra
terms to remove the constant and linear in $a$ terms in \eq{R23} we
obtain our final answer for $R_2$:
\begin{align}\label{R25}
  R_2 \, \equiv \, \frac{i}{2} \, \frac{192^2}{Q^{12}} \,
  \int\limits_0^{\infty} \, \frac{d \, y}{(1+y)^{12}} \, y \, (1-y)^2
  \, \left[ 1 + i \, q^+ \, a \, (1 + y) - e^{i \, q^+ \, a \, (1 +
      y)} \right]
\end{align}


We now need to evaluate 
\begin{align}\label{RL21}
  R_2^L \, = & \, \int\limits_{-\infty}^\infty \frac{d \, \xi}{2 \, \pi}
  \, \frac{1}{(1-\xi + i \, \epsilon)^2} \, \left[ 2 - e^{i \, q^+ \,
      a \, (1 - \xi + i \, \epsilon)} - e^{- i \, q^+ \, a \, (1 - \xi
      + i \, \epsilon)} \right] \notag \\ & \times \,
  \int\limits_0^\infty \, d z \, z^5 \, K_0 (z \, Q) \,
  \int\limits_0^\infty d z' \, I_0 \left( z_< \, Q \, \sqrt{\xi - i \,
      \epsilon} \right) \, K_0 \left( z_> \, Q \, \sqrt{\xi - i \,
      \epsilon} \right) \, z'^5 \, \, K_0 (z' \, Q).
\end{align}
needed for calculation of the longitudinal components of the hadronic
tensor in \eq{2-term2++}. We begin by employing the series
representation for modified Bessel functions $I_0$ and $K_0$
\begin{subequations}\label{IK0ser}
\begin{align}
  I_0 (z) \, & = \, \sum\limits_{m=0}^\infty \, \frac{1}{[\Gamma
    (m+1)]^2} \, \left( \frac{z}{2} \right)^{2 \, m}  \\
  K_0 (z) \, & = \, - \sum\limits_{m=0}^\infty \, \frac{1}{[\Gamma
    (m+1)]^2} \, \left( \frac{z}{2} \right)^{2 \, m} \, \left[ \ln
    \left( \frac{z}{2} \right) - \psi (m+1) \right]
\end{align}
\end{subequations}
to infer that the complex $\xi$-plane structure of the integrand in
\eq{RL21} is the same as shown in \fig{compl} above. Picking up the
pole at $\xi = 1 + i \, \epsilon$ and wrapping the contour around the
branch cut yields, similar to \eq{R22},
\begin{align}\label{RL22}
  R_2^L \, = & \, \frac{i}{2} \int\limits_{-\infty}^0 \, \frac{d \,
    \xi}{(1-\xi)^2} \, \left[ 2 - e^{i \, q^+ \, a \, (1 - \xi)}
  \right] \left[ \int\limits_0^\infty \, d z \, z^5 \, K_0 (z \, Q) \,
    J_0 \left( z \, Q \, \sqrt{-\xi} \right) \right]^2 + i \,
  \frac{d}{d \xi} \Bigg\{ \left[ 2 - e^{i \, q^+ \, a \,
      (1 - \xi)} \right] \notag \\
  & \times \, \int\limits_0^\infty \, d z \, z^5 \, K_0 (z \, Q) \,
  \int\limits_0^\infty d z' \, I_0 \left( z_< \, Q \, \sqrt{\xi}
  \right) \, K_0 \left( z_> \, Q \, \sqrt{\xi} \right) \, z'^5 \, \,
  K_0 (z' \, Q) \Bigg\} \Bigg|_{\xi = 1}
\end{align}
where we have used
\begin{align}
  I_0 \left( z_< \, Q \, \sqrt{\xi - i \, \epsilon} \right) \, K_0
  \left( z_> \, Q \, \sqrt{\xi - i \, \epsilon} \right) - I_0 \left(
    z_< \, Q \, \sqrt{\xi + i \, \epsilon} \right) \, K_0 \left( z_>
    \, Q \, \sqrt{\xi + i \, \epsilon} \right) \notag \\ \, = \, \pi
  \, i \, \theta (-\xi) \, J_0 \left( z \, Q \, \sqrt{-\xi} \right) \,
  J_0 \left( z' \, Q \, \sqrt{-\xi} \right)
\end{align}
which follows from Eqs.~(\ref{IK0ser}). The second term on the
right-hand-side in \eq{RL22} is again a linear polynomial in $a$, with
the coefficients that can be fixed by requiring that the Taylor
expansion of $R_2^L$ in powers of $a$ starts from order-$a^2$.
Imposing this condition and integrating over $z$ in the first term on
the right-hand-side of \eq{RL22} we arrive at the final result
\begin{align}\label{RL23}
  R_2^L \, = \, \frac{i}{2} \, \frac{64^2}{Q^{12}} \,
  \int\limits_0^{\infty} \, \frac{d \, y}{(1+y)^{12}} \, (1 - 4 \, y +
  y^2)^2 \, \left[ 1 + i \, q^+ \, a \, (1 + y) - e^{i \, q^+ \, a \,
      (1 + y)} \right]
\end{align}
where, as before, $y = - \xi$.



\providecommand{\href}[2]{#2}\begingroup\raggedright\endgroup


\end{document}